\documentclass[12,preprint]{aastex}
\usepackage{epsfig}
\usepackage{graphics}
\usepackage{epstopdf}
\usepackage{amsmath}
\usepackage{color}
\usepackage{lscape}

\newcommand{\irf}[1]{\texttt{#1}}

\newlength{\twothirdscolfigwidth}
\usepackage{cite}
\usepackage{natbib}
\usepackage{aas_macros}
\slugcomment{}
\begin{document}

\title{Determination of the Point-Spread Function for the {\it Fermi} Large Area Telescope from On-orbit Data and Limits on Pair Halos of Active Galactic Nuclei}

\begin{abstract}
  The Large Area Telescope (LAT) on the \textit{Fermi} Gamma-ray Space
  Telescope is a pair-conversion telescope designed to detect photons
  with energies from $\approx$20 MeV to $> $300 GeV.
  The pre-launch response functions of the LAT
  were determined through extensive Monte Carlo simulations and beam
  tests. 
  The point-spread function (PSF) characterizing the angular
  distribution of reconstructed photons as a function of energy and
  geometry in the detector is determined here from two years of
  on-orbit data by examining the distributions of $\gamma$ rays from
  pulsars and active galactic nuclei (AGN).
  Above 3 GeV, the PSF is found to be broader than the pre-launch PSF.
  We checked for dependence of the PSF on the class of $\gamma$-ray
  source and observation epoch and found none.  We also investigated
  several possible spatial models for pair-halo emission around BL Lac
  AGN.  We found no evidence for a component with spatial extension
  larger than the PSF and set upper limits on the amplitude of halo
  emission in stacked images of low and high redshift BL Lac AGN and
  the TeV blazars 1ES0229$+$200 and 1ES0347$-$121.
\end{abstract}

\pagestyle{empty}
\author{
M.~Ackermann\altaffilmark{1}, 
M.~Ajello\altaffilmark{2}, 
A.~Allafort\altaffilmark{2}, 
K.~Asano\altaffilmark{3}, 
W.~B.~Atwood\altaffilmark{4}, 
L.~Baldini\altaffilmark{5}, 
J.~Ballet\altaffilmark{6}, 
G.~Barbiellini\altaffilmark{7,8}, 
D.~Bastieri\altaffilmark{9,10}, 
K.~Bechtol\altaffilmark{2}
% ...
}
\altaffiltext{1}{Deutsches Elektronen Synchrotron DESY, D-15738 Zeuthen, Germany}
\altaffiltext{2}{W. W. Hansen Experimental Physics Laboratory, Kavli Institute for Particle Astrophysics and Cosmology, Department of Physics and SLAC National Accelerator Laboratory, Stanford University, Stanford, CA 94305, USA}
\altaffiltext{3}{Interactive Research Center of Science, Tokyo Institute of Technology, Meguro City, Tokyo 152-8551, Japan}
\altaffiltext{4}{Santa Cruz Institute for Particle Physics, Department of Physics and Department of Astronomy and Astrophysics, University of California at Santa Cruz, Santa Cruz, CA 95064, USA}
\altaffiltext{5}{Istituto Nazionale di Fisica Nucleare, Sezione di Pisa, I-56127 Pisa, Italy}
\altaffiltext{6}{Laboratoire AIM, CEA-IRFU/CNRS/Universit\'e Paris Diderot, Service d'Astrophysique, CEA Saclay, 91191 Gif sur Yvette, France}
\altaffiltext{7}{Istituto Nazionale di Fisica Nucleare, Sezione di Trieste, I-34127 Trieste, Italy}
\altaffiltext{8}{Dipartimento di Fisica, Universit\`a di Trieste, I-34127 Trieste, Italy}
\altaffiltext{9}{Istituto Nazionale di Fisica Nucleare, Sezione di Padova, I-35131 Padova, Italy}
\altaffiltext{10}{Dipartimento di Fisica ``G. Galilei", Universit\`a di Padova, I-35131 Padova, Italy}
% ...

\maketitle

\begin{doublespace}

\section{Introduction}

The Large Area Telescope (LAT) is the primary instrument on board the {\it Fermi Gamma-ray Space Telescope}, launched in 2008, and is sensitive to $\gamma$ rays from $\approx 20$ MeV to $> 300$ GeV \citep{REF:2009.LATPaper}. 
The LAT consists of a $4\times 4$ array of modules called \textit{towers}, each with tracker and calorimeter sections, surrounded by a segmented anticoincidence detector to veto charged particles. 
The tracker sections have 18 layers of alternating x-y pairs of silicon strip detectors.
Each of the first 16 layers of the tracker has a layer of tungsten foil to induce $\gamma$ rays to pair convert.
Pair conversion in tungsten layers in the LAT tracker creates secondary $e^{+}$--$e^{-}$ pairs that deposit ionization energy in the silicon tracker layers. 
The direction of the original $\gamma$ ray is reconstructed from the tracks of the secondaries, and the energy of the $\gamma$ ray is determined from the energy deposition in the calorimeter and the estimated energy losses in the tracker. 
See \citet{REF:2007.TKRPaper} for further details of the tracker system.

At energies below $\approx 10$ GeV, the accuracy of the directional reconstruction is limited by multiple scattering, whereas above $\approx 10$ GeV, the accuracy is limited by the lever arm of the direction measurement and the 228-$\mu$m silicon-strip pitch. 
By design, the tracker has significantly different angular resolution depending on whether the incident $\gamma$ ray converts in the \textit{front} or the \textit{back} tungsten layers. 
The twelve front conversion planes have thinner tungsten layers (3\% of a radiation length) and longer track lengths for the converted pairs, yielding good angular resolution but smaller conversion efficiency. 
The four back conversion planes have thicker layers of tungsten (18\% of a radiation length) to increase the effective area and field of view at high energies, but provide poorer angular resolution due to the increased multiple scattering and shorter track lengths. 

The point-spread function (PSF) of the LAT is the probability
distribution function (PDF) $p(\delta v;E,\hat{v})$, for $\delta v = |\hat{v}-\hat{v}'|$, the offset
between the true($\hat{v}$) and reconstructed($\hat{v}'$) directions of the
$\gamma$ ray of true energy $E$.
The characteristic angular size of the PSF scales with energy as the sum of the angular
uncertainties due to the instrument-pitch and multiple scattering, added in quadrature.  We parameterize this energy
dependence with the scaling function,
\begin{equation}
  S_{P}(E)\propto
\sqrt{\left(c_{0}~\left(\frac{E}{100\textrm{ MeV}}\right)^{-\beta}\right)^{2} + c_{1}^2}\;,
\label{eq:psffront}
\end{equation}
where $c_{0}$ is the normalization of the multiple scattering term,
$c_{1}$ is instrument-pitch uncertainty, and $\beta$ sets the scaling
of the multiple scattering with energy.  The 68\% containment radius
for front-converting events can be approximated with Equation
\ref{eq:psffront} and $c_{0}$ = 3.5$^{\circ}$, $c_{1}$ =
0.15$^{\circ}$, and $\beta \approx 0.8$ \citep{REF:2009.LATPaper}.
The angular resolution for a $\gamma$ ray converting in the
back layers is typically about a factor of two larger than
for the front layers.

Accurate characterization of the LAT PSF is critical for proper source
analysis.  It has assumed additional importance because of the
potential for inferring the magnitude of the intergalactic magnetic
field $B_{IGMF}$ from the measurement of $\gamma$-ray halos around TeV
blazars, a subset of active galactic nuclei (AGN), e.g.,
\citet{2009PhRvD..80b3010E}.  In intergalactic space, $B_{IGMF}$ could
be a remnant of exotic processes taking place in the early universe,
far earlier than the decoupling epoch \citep{2009PhRvD..80l3012N}.
TeV $\gamma$ rays annihilating due to $\gamma$-$\gamma$ interactions
with the extragalactic background light (EBL) create relativistic
electrons and positrons that Compton-scatter photons of the cosmic
microwave background (CMB) to GeV energies.  Depending on the
magnitude and correlation length of $B_{IGMF}$, a halo of secondary
GeV photons with a characteristic spectrum
\citep{2007A&A...469..857D,REF:2010.Neronov2} and angular extent
\citep{2009PhRvD..80b3010E} will surround a TeV source.

The MAGIC collaboration examined several potential angular profiles of
emission for the blazars Mrk 421 and 501 at energies
between 0.3 -- 3.0 TeV and found no significant extension compared to
their PSF width of $0.1^{\circ}$.  
The upper limit on the flux of extended emission from Mrk 501 may
constrain magnetic field strengths in the range $4 \times 10^{-15} -
1.3 \times 10^{-14} \textrm{ G}$ if the IGMF coherence length scale is
$\gtrsim 1$~Mpc
\citep{REF:2010.MAGIC}.  \citet{REF:2010.AndoKus} have recently
claimed that halos around bright AGN in LAT data give direct evidence
for $B_{IGMF}\approx 10^{-15}$ G.  A subsequent analysis by
\citet{REF:2011.Neronov1} of the LAT data found no significant
pair-halo component of AGN when compared with the profile of the Crab
pulsar and nebula.  A detailed investigation of the {\it Fermi}-LAT
PSF is performed in this paper using the Vela and Geminga pulsars and
AGN.


The functional representation used to characterize the PSF is described in Section 2. 
In Section 3, the on-orbit PSF derived from an analysis of AGN is compared with the PSF inferred from simulations and cross-checked against an analysis of pulsars. 
We also consider possible contributions from systematic effects for the measured differences between the pre-launch and the on-orbit PSF. 
In Section 4, we present an analysis that quantifies the limits on halo emission around AGN.
Discussion and summary are given in Section 5.

\section{Instrument Response and the Point-Spread Function}

To parameterize the instrument response of the LAT, we performed Monte Carlo (MC) simulations of large samples of $\gamma$ rays and charged particles, and analyzed these data with our event reconstruction and classification algorithms \citep{REF:2009.LATPaper}. 
The physical interaction of particles with the LAT is modeled by an implementation of the Gaudi\footnote{Gaudi: Gaudi Project http://proj-gaudi.web.cern.ch/proj-gaudi/} framework
 and the Geant4 toolkit \citep{2003NIMPA.506..250G}, Gleam \citep{REF:2003sngh.conf..141B}. 
We generate a high-statistics, uniform $\gamma$-ray dataset using Gleam, and we apply to the events a simulation of the LAT trigger and on-board filter. 
Accepted events are reconstructed and undergo the same event analysis scheme as real events, the details of which may be found in \citet{2012ApJS..203....4A}. 
The simulated particles that survive triggering and filtering are then passed through the event reconstruction and classification.
The resulting MC sample was used to determine the pre-launch effective area, energy dispersion, and PSF as a function of energy and inclination angle, or the angle from the boresight \citep{REF:2009.LATPaper}. 
A calibration unit consisting of two tracker modules and three calorimeter modules was tested with charged-particle and photon beams to 
validate the MC simulation \citep{REF:2007.TKRPaper}. 

\begin{figure}[htb]
\centering
\includegraphics[scale=0.7]{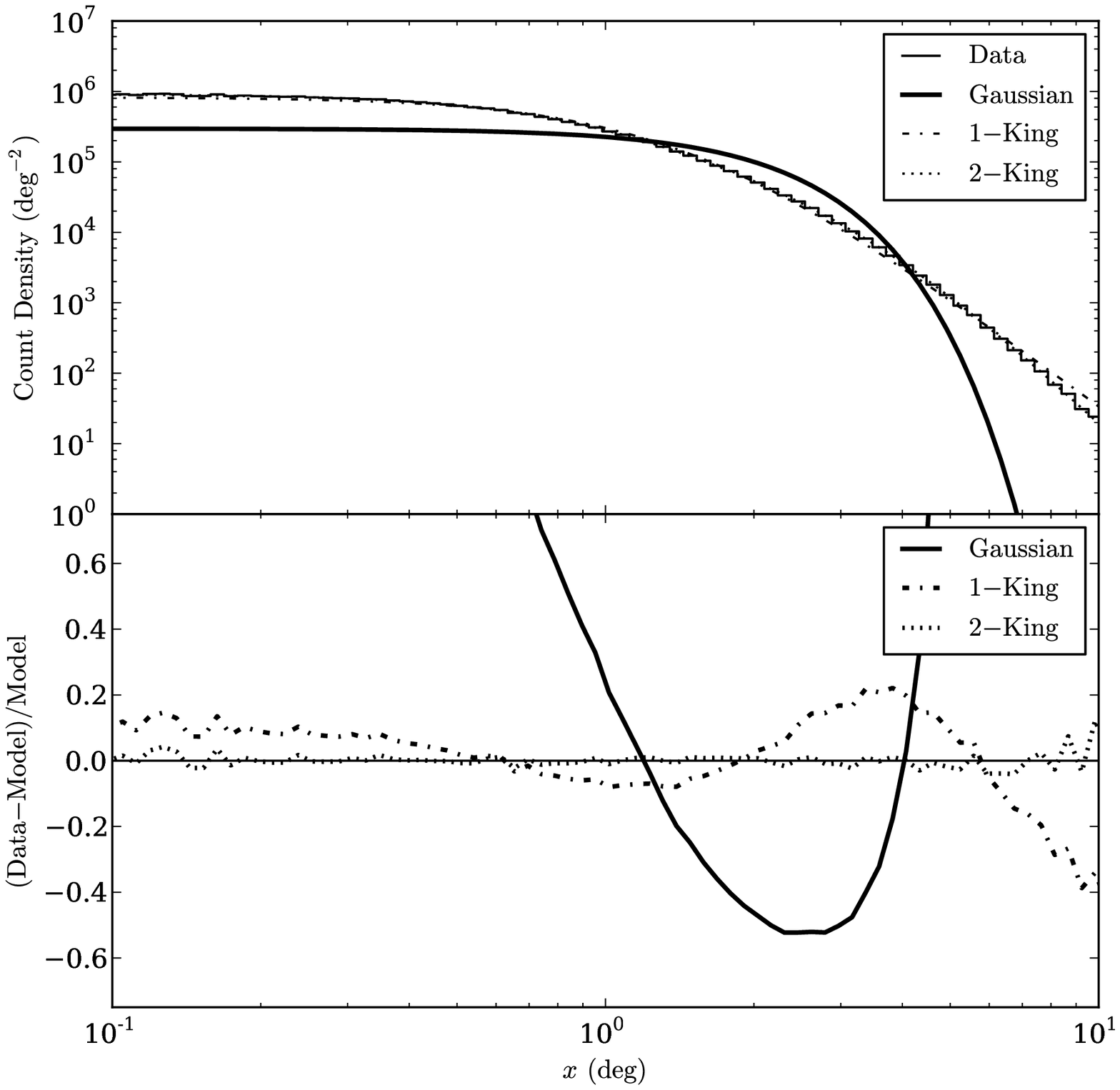}
\caption{Comparison of best-fit Gaussian, single, and double King function fits to the angular distribution of simulated \textit{Diffuse}-class $\gamma$ rays with energy 7.5 GeV impinging at inclination angles between 26 and 37$^\circ$ uniformly in solid angle. The best-fit Gaussian, determined by binned likelihood, gives a poor representation of the PSF at small and large separations because of power-law tails at large angles. 
}
\label{FIGURE::PSFDIFF}
\end{figure}

We assume azimuthal symmetry of the PSF with respect to the
$\gamma$-ray direction such that the PSF can be described by a PDF
with a single parameter $\delta v = |\hat{v}'-\hat{v}|$, the angular
devation between the reconstructed and true $\gamma$-ray direction.
Furthermore, we assume that the PSF does not depend on the azimuth
angle of the incoming $\gamma$ ray but
only on the inclination angle with respect to the LAT boresight, $\theta$.  Thus the PSF can be written
as $P(\delta v; E, \theta)$.  Since the angular size of the PSF varies
by 2 orders of magnitude over the LAT energy range, we use Equation
\ref{eq:psffront} to scale out most of the energy dependence of the
PSF by expressing it as a PDF in the \textit{scaled angular deviation},
\begin{equation}
x=\frac{\delta v}{S_{P}(E)}.
\end{equation}
We parameterize the PSF in terms of the King function \citep[cf.][]{1962AJ.....67..471K}, which was chosen to follow the power-law behavior of the PSF at large angular offsets. 
The King function has the form
\begin{equation}
K(x,\sigma,\gamma) =\frac{1}{2\pi\sigma^{2}} \left(1 - \frac{1}{\gamma}\right)\left(1+\frac{x^{2}}{2\sigma^{2}\gamma}\right)^{-\gamma},
\label{eq:king}
\end{equation}
and $K$ is normalized in the small-angle approximation $d\Omega= x dx d\phi$, so that 
\begin{equation}
\int_0^{2\pi} \int_0^{\infty} K(x,\sigma,\gamma)xdxd\phi = 1.
\end{equation}
The parameter $\sigma$ is the characteristic size of the angular distribution and $\gamma$ determines the weight of the tails of the distribution. 
The King function becomes the normal distribution in the limit $\gamma\rightarrow\infty$, where $1.51\sigma\approx68\%$ containment. 
A best-fit single King function fit is not enough to properly reproduce the observed distributions of simulated $\gamma$ rays (see Figure \ref{FIGURE::PSFDIFF}), so we opted for a more complex model using the sum of two King functions
\begin{equation}
\begin{split}
P(x,E) = f_{core} K(x,\sigma_{core}(E),\gamma_{core}(E)) 
+ (1-f_{core}) K(x,\sigma_{tail}(E),\gamma_{tail}(E))\;
\end{split}
\label{eq:psfform}
\end{equation}
where the \textit{core} and \textit{tail} components characterize the distributions for small and large angular separations, respectively.
As can be seen from Figure \ref{FIGURE::PSFDIFF}, this functional form provides a good fit to the simulated angular distributions of event counts 
around the $\gamma$-ray direction. 
The parameters $\gamma_{core}$ and $\gamma_{tail}$ determine the structure of the PSF tail and are found from simulations to decrease at high energy, 
yielding larger tail fractions above $\approx$ 10 GeV. 
For the analysis described in Section 3, we also use, as an alternative to the King function, a model independent form of the PSF 
parameterized by the angles corresponding 
to the 68 and 95\% integral fractions of the distribution, or the 68 and 95\% containment radii. 

The LAT Science Tools distribution has for each event class a set of tables that contain the PSF parameters for each conversion type (front or back) as a function of energy and inclination angle\footnote{{\it Fermi} LAT Science Tools are found at http://fermi.gsfc.nasa.gov/ssc/data/analysis/documentation/}.
By scaling out the energy dependence of the angular size $\sigma$, the tables isolate the energy dependence of the PSF tails and the weak dependence on the inclination angle. 
The parameters are tabulated for logarithmic energy intervals ranging from 18 MeV to 562 GeV with 4 bins per decade, and in the cosine of the inclination angle between 0.2 and 1.0 in increments of 0.1. 
The tables contain all of the parameters described in Equation \ref{eq:psfform} and the scaling function for the energy dependence of the $\sigma$ parameters, which has the form given by Equation \ref{eq:psffront}.

The PSF parameters determined from MC simulations for the 
pre-launch event analysis (\irf{Pass6}) were the first publicly released set \citep[e.g.,][]{REF:2009.LATPerf}, denoted as \irf{Pass6} version 3 (\irf{P6\_V3}). 
A range of event classifications for LAT data is available, to meet different scientific requirements. 
The \textit{Diffuse} event class selection has the highest quality requirements on track reconstruction and low charged-particle background and is most suitable for analyzing weak point sources. 
For these reasons, we derive the PSF for the \textit{Diffuse} event class only.

Since the release of the \irf{Pass6}, a newer version of the event analysis,
\irf{Pass7}, has been publicly released that integrates all of the known on-orbit effects
into the instrument response.
The PSFs associated with each of the \irf{Pass7} event classes suitable for source analysis were derived using the methodology outlined in Section 3, identical to the on-orbit PSF analysis of \irf{Pass6} \textit{Diffuse} event class data.
The PSFs of these event classes are described and validated in
\citet{2012ApJS..203....4A}.


\section{On-orbit PSF}

The primary motivation for determining the PSF from LAT data was to verify that the reconstruction of $\gamma$ rays was consistent with the simulations described in Section 2.
When performing analyses of point sources using months of accumulated statistics, we noticed discrepancies between the PSF derived from MC simulations and the angular distributions of measured directions of $\gamma$ rays around bright point sources, with the observed distributions systematically broader for energies above a few GeV. 
Over time, sufficient $\gamma$-ray statistics accumulated for an accurate characterization of the on-orbit PSF at high energies. 
The leading considerations in the development of the on-orbit PSF for \irf{Pass6} data (\irf{P6\_V11}) were twofold: 
1) reproducing the angular profiles of point sources in the LAT data, thereby limiting the biases and systematic uncertainties in measurements of spatial extensions and spectra of sources, and 
2) smoothing the energy dependence of the PSF parameterization to avoid introducing spurious features from statistical fluctuations that could affect the quality of source analysis. 

In this section 
we describe the stacking analysis used to determine the on-orbit PSF and the validation of the methodology using a simulation. We then verify the on-orbit PSF derived from this analysis using pulsars and AGN.
Finally, we evaluate sources of systematic uncertainties in the calibration of the instrument that may influence the PSF.

\subsection{PSF Derived from Stacked AGN}\label{SECTION::ONORBITANALYSIS}
To calibrate the PSF, we adopted a technique of stacking sources,
where the angular offsets of $\gamma$ rays from their presumed sources
are analyzed as if they came from a single source.  Pulsars would be
ideal for calibrating the PSF. However, the $\gamma$ rays from pulsars
above 10 GeV are limited, so we restrict our analysis to AGN.
A subset of 65 AGN was selected from the \textit{Fermi} Large Area
Telescope First Source Catalog, \citep[henceforth
1FGL,][]{REF:2010.1FGL} to create a calibration sample.  All the AGN
in the sample have flux between 1$-$100 GeV of at least 1.66 $\times$
10$^{-9}$ photons cm$^{-2}$ s$^{-1}$ and a
\textit{TS}\footnote{\textit{TS}: Test Statistic, 2$\Delta$log(likelihood) between
  models with and without the source, c.f. \citet{1996ApJ...461..396M}} of at least 81 above 1 GeV; a
list of the sources and their properties is given in Table
\ref{TABLE::PSFAGN}.  Out of the 65 sources, 35 were associated with
BL Lac-type blazars, 27 with Flat Spectrum Radio Quasars (FSRQ-type blazars), 
1 with a non-blazar active galaxy, and 2 with
an active galaxy of uncertain type.
Though not an explicit criterion for selecting the AGN sample, the
calibration sample of bright AGN are primarily at high Galactic
latitude $|b|>10^\circ$, limiting the possible systematic
uncertainties associated with the background intensity and structure
of the Galactic diffuse emission.


For our analysis, we used the events in the \irf{P6\_V3}
\textit{Diffuse} class of $\gamma$ rays from the 24-month period 2008
August 4 $-$ 2010 August 4 (mission elapsed time in the range
239557417 s $-$ 302629417 s).  We selected events with energies
between 1 and 100 GeV in a region of interest (ROI) of radius
$4^{\circ}$ around each source. The lower energy limit was chosen to
limit contamination in the ROIs from nearby point sources, since for
the average source separation $\approx 7.0^\circ$ for our sources
above a Galactic latitude $|b|>10^\circ$, the 99\% containment radii
begin to overlap at about this energy.  We also excluded events with
zenith angle greater than $105^{\circ}$ in order to limit the
contamination from the Earth limb. We excluded inclination angle in
the detector greater than $66.4^{\circ}$ to remove $\gamma$ rays for
which the PSF is significantly broader and the acceptance is small
($\lesssim 0.2\%$ of the total acceptance).  The data were split into
their respective conversion types, front and back, and into four
energy bins per logarithmic decade between 1 and 32 GeV and a single
bin from 32 to 100 GeV.  Due to the limited statistics in the energy
bins above 10 GeV, the events were not binned in inclination
angle. 
The model we derive from this analysis is representative of the PSF
over exposures longer than the 53.4 day orbital precession period of
the spacecraft.  The PSF for a given source and observation period
depends on the source observing profile, the accumulated exposure of
the source as a function of its inclination angle in the LAT.  Over
long exposures the observing profile converges to an approximately
constant shape and the use of an average PSF model is well justified.
On shorter time scales, the shape of the observing profile can differ
significantly and potentially introduce significant variations in the
PSF relative to the one derived in this analysis.


We chose to model the angular distribution of $\gamma$ rays around our
sample of AGN for a given energy bin as the sum of a single King
function and an isotropic background from diffuse $\gamma$-ray
emission and residual cosmic rays.
The single King function parameterization was chosen over the
double-King function used to model the MC PSF in order to stabilize
the convergence of the PSF parameters when fitting to the lower
statistics of the stacked AGN sample.
The log-likelihood $\log L$ for the model of the stacked counts distribution for each energy bin given the observations is defined as
\begin{equation}
\log L(N_{psf}, N_{iso}, \sigma, \gamma \mid \vec{x}) = -N_{psf} -N_{iso} + \sum_{j=1}^N \textrm{log} \left( N_{psf} K(x_j,\sigma,\gamma) + N_{iso}~ I\right)\;,
\label{eq:loglike}
\end{equation}
where $K$ is the normalized King function from Equation \ref{eq:king}
as a function of angular separation $x_j$, $I$ is the isotropic
normalization factor, and $j$ is summed over the number of $\gamma$
rays $N$.
The localization uncertainty in the 1FGL positions was orders of magnitude smaller than the PSF, so we used the measured positions as the reference directions to determine the $\gamma$-ray angular separations. 

\begin{figure}[htb]
\centering
\includegraphics[scale=0.43]{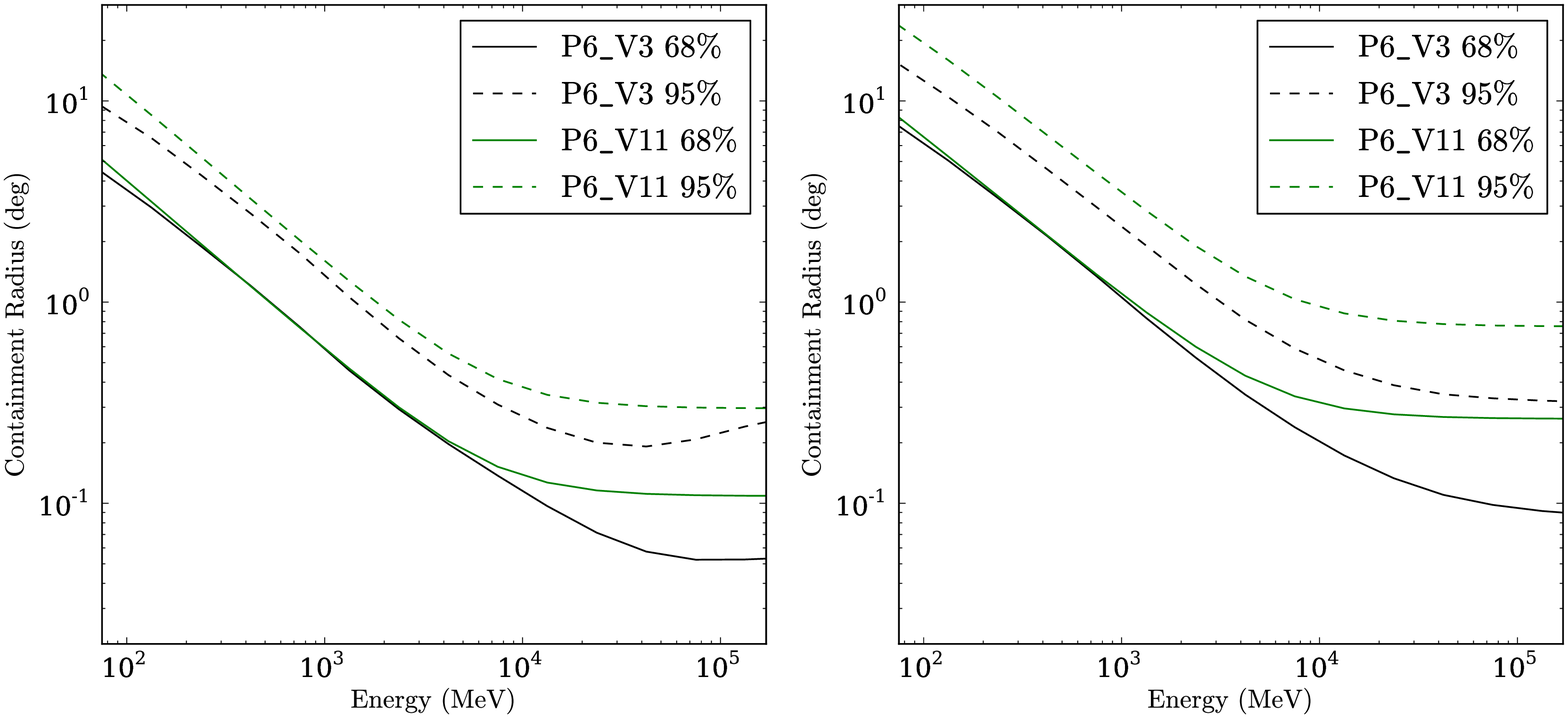}
\caption{Comparison of 68\% (solid lines) and 95\% (dashed lines) containment
radii of the MC (black) and on-orbit (green) PSF models for front- (left) and
back-converting (right) \textit{Diffuse}-class events.}
\label{FIGURE::PSFCOMP}
\end{figure}


While the King function parameter $\sigma$ of the PSF has the same
scaling with energy as the characteristic angular size of Equation
\ref{eq:psffront}, the tail parameter $\gamma$ has a more complicated
energy dependence.  
Because small changes in $\gamma$ can induce large changes in the
shape of the PSF, we chose to reparameterize the King function in
terms of the 68\% and 95\% containment radii, $R_{68}$ and $R_{95}$.
For a single King function (see Equation \ref{eq:king}) any two
containment radii uniquely determine the parameters $\sigma$ and
$\gamma$.  While $\sigma$ and $\gamma$ cannot be expressed
analytically in terms of the two containment radii, they can
nonetheless be determined numerically and we denote them as
$\sigma(R_{68},R_{95})$ and $\gamma(R_{68},R_{95})$.  We model the
energy-dependence of the 68\% and 95\% containment radii with two
independent scaling functions, $R_{68}(E)$ and $R_{95}(E)$, with the
form given by Equation \ref{eq:psffront} and each with three
independent parameters: $\mathbf{c} = \{c_0,c_1\}$ and $\beta$.

We used a maximum likelihood analysis to determine the best fit
parameters for $R_{68}(E)$ and $R_{95}(E)$ by maximizing the sum of
the log-likelihoods from Equation \ref{eq:loglike} over all the energy
bands.  The joint log-likelihood $\log L$ is defined as
\begin{equation}
\begin{split}
  \log L(\mathbf{c}_{68},\beta_{68},&
  \mathbf{c}_{95},\beta_{95} \mid 
\{\vec{x}_{0}, \ldots, \vec{x}_M\}) = \\
  &\sum_i^{M} \log L (N_{psf}^i,
  N_{iso}^i,\sigma(R_{68}(E_{i}),R_{95}(E_{i})),
  \gamma(R_{68}(E_{i}),R_{95}(E_{i})) \mid \vec{x}_i),
\end{split}
\label{eq:joint}
\end{equation}
where $\log L$ is the log-likelihood from Equation \ref{eq:loglike},
$\vec{x}_{i}$ is the set of angular separations in energy bin $i$,
$E_i$ is the bin energy, and the parameter dependence of $R_{68}(E)$
and $R_{95}(E)$ is implied.  Given the best fit scaling functions, we
extract the King function parameters for the on-orbit PSF tables by
evaluating $\sigma(R_{68}(E),R_{95}(E))$ and
$\gamma(R_{68}(E),R_{95}(E))$ at the geometric mean energy of each
bin.  This procedure creates a set of PSF parameters that are smoothly
varying with energy.
The resulting parameterization is an extrapolation of the PSF outside
the energy range of the analysis: below 1 GeV and above 100 GeV.
Figure \ref{FIGURE::PSFCOMP} compares the MC (\irf{P6\_V3}) and
on-orbit (\irf{P6\_V11}) PSFs.

\begin{figure}[htb]
\centering
\includegraphics[scale=0.43]{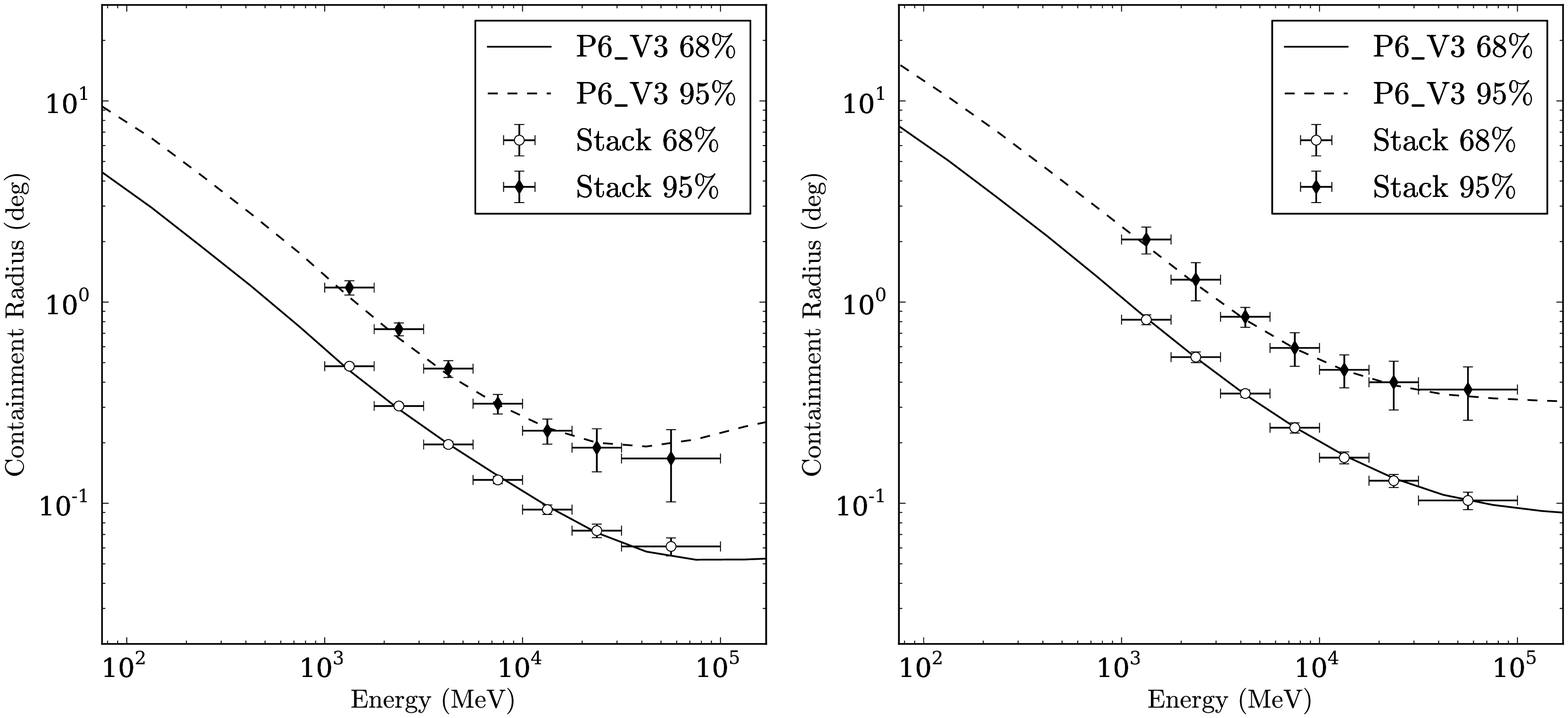}
\caption{68\% (open circles) and 95\% (black diamonds) PSF containment radii
inferred from applying the on-orbit PSF analysis to a simulated AGN
sample (Stack) with the properties of the calibration AGN sample
generated with the \irf{P6\_V3} \textit{Diffuse} PSF for front- (left) and back-converting
(right) events. Solid lines show the containment radii predicted by
the same PSF used to generate the simulation.}
\label{FIGURE::MCSTACK}
\end{figure}

To examine the potential bias of determining the PSF from on-orbit
data, specifically the extrapolation of the PSF beyond the 1-100 GeV
range, we generated and analyzed a detailed simulation of the sky.  We
simulated all sources from the \textit{Fermi} Large Area Telescope
Second Source
Catalog 
\citep[2FGL;][]{2012ApJS..199...31N}, along with the Galactic and
isotropic diffuse models, using the Science Tool \textit{gtobssim}
with the \irf{P6\_V3} \textit{Diffuse} PSF.  The simulation covered
the same time span as the on-orbit data selection and used the same
cuts on inclination angle, energy, and zenith angle.  We chose to use
the 2FGL catalog for the simulation to account for the presence of
sources not in the 1FGL that could introduce structured background and
create a systematic uncertainty in the PSF
determination. 
We analyzed the simulation with the same set of 65 AGN from the
on-orbit PSF analysis and determined the simulation PSF in the same
manner as the on-orbit data, using Equation \ref{eq:joint}.  The 68\%
and 95\% containment radii determined by the PSF analysis of the
simulated data are compared with those derived numerically from the
\irf{P6\_V3} PSF in Figure \ref{FIGURE::MCSTACK}.  The \irf{P6\_V3}
containment radii were derived by averaging the PSF model over
inclination angle weighted by the effective area.  We find good
agreement between the 68\% and 95\% containment radii of the
\irf{P6\_V3} PSF and the containment radii derived from the sky
simulation.  Additionally, we find the containment radii extrapolated
below 1 GeV and above 100 GeV are in good agreement with the measured
values at these energies.  This finding is consistent with the
expectation from Equation \ref{eq:psffront} that the containment
should follow the $E^{-\beta}$ scaling from muliple scattering below 1
GeV and take a constant value above 100 GeV.


We determined that there were no large systematic uncertainties from the determination of the PSF by the stacking technique. 
However, we sought to verify the on-orbit PSF using a different technique and class of point sources. 
The Vela and Geminga pulsars are the brightest persistent $\gamma$-ray point sources in the 100 MeV$-$10 GeV energy range and are alternate calibration sources for the PSF. 
In the next section, we use these pulsars to cross-check against the PSF inferred from AGN, including the extrapolation below 1 GeV. 

\subsection{On-orbit PSF Verification}\label{SECTION::ONORBITVALID}

\begin{figure}[htb]
\centering
\includegraphics[scale=0.86]{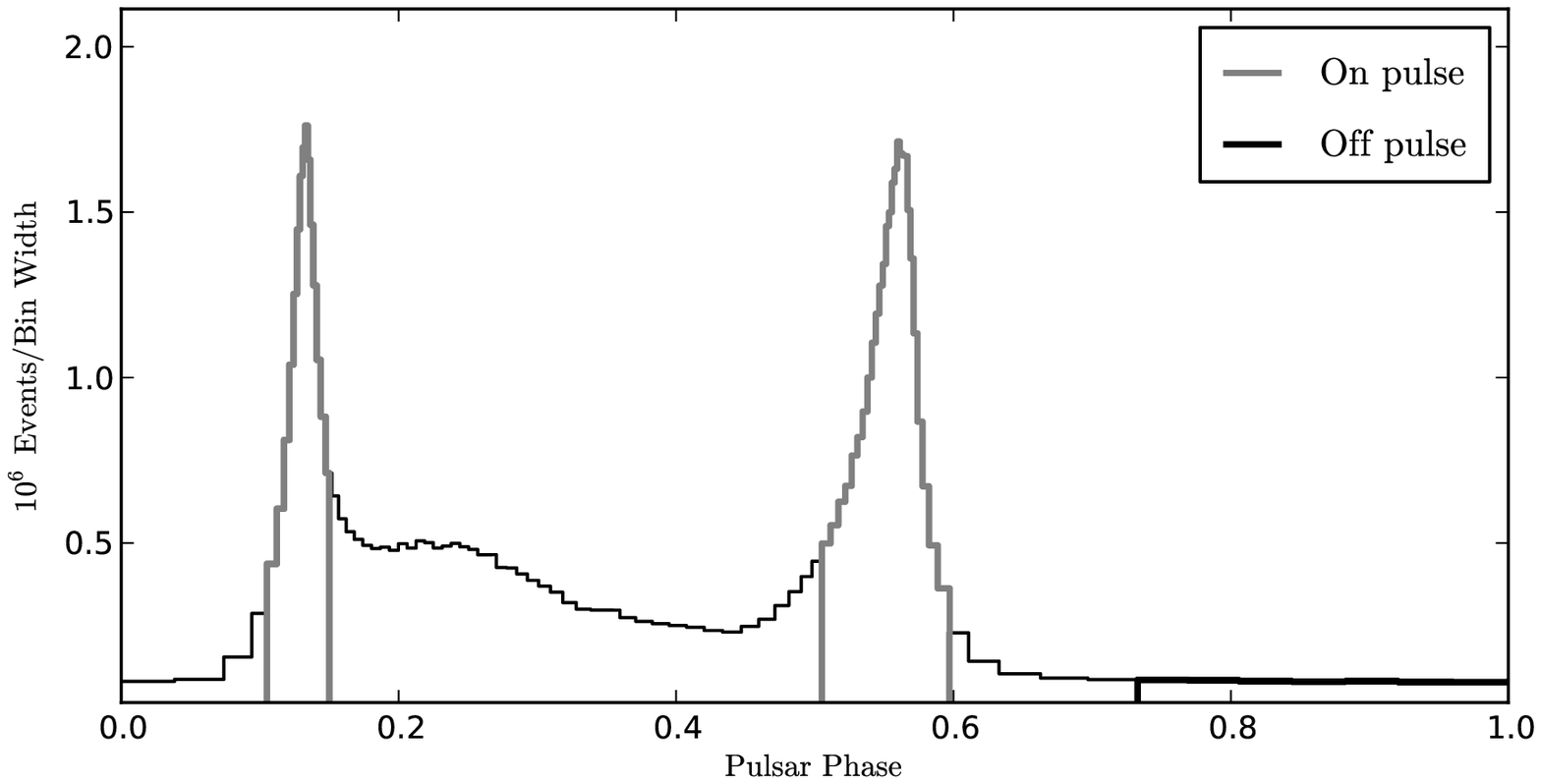}
\includegraphics[scale=0.86]{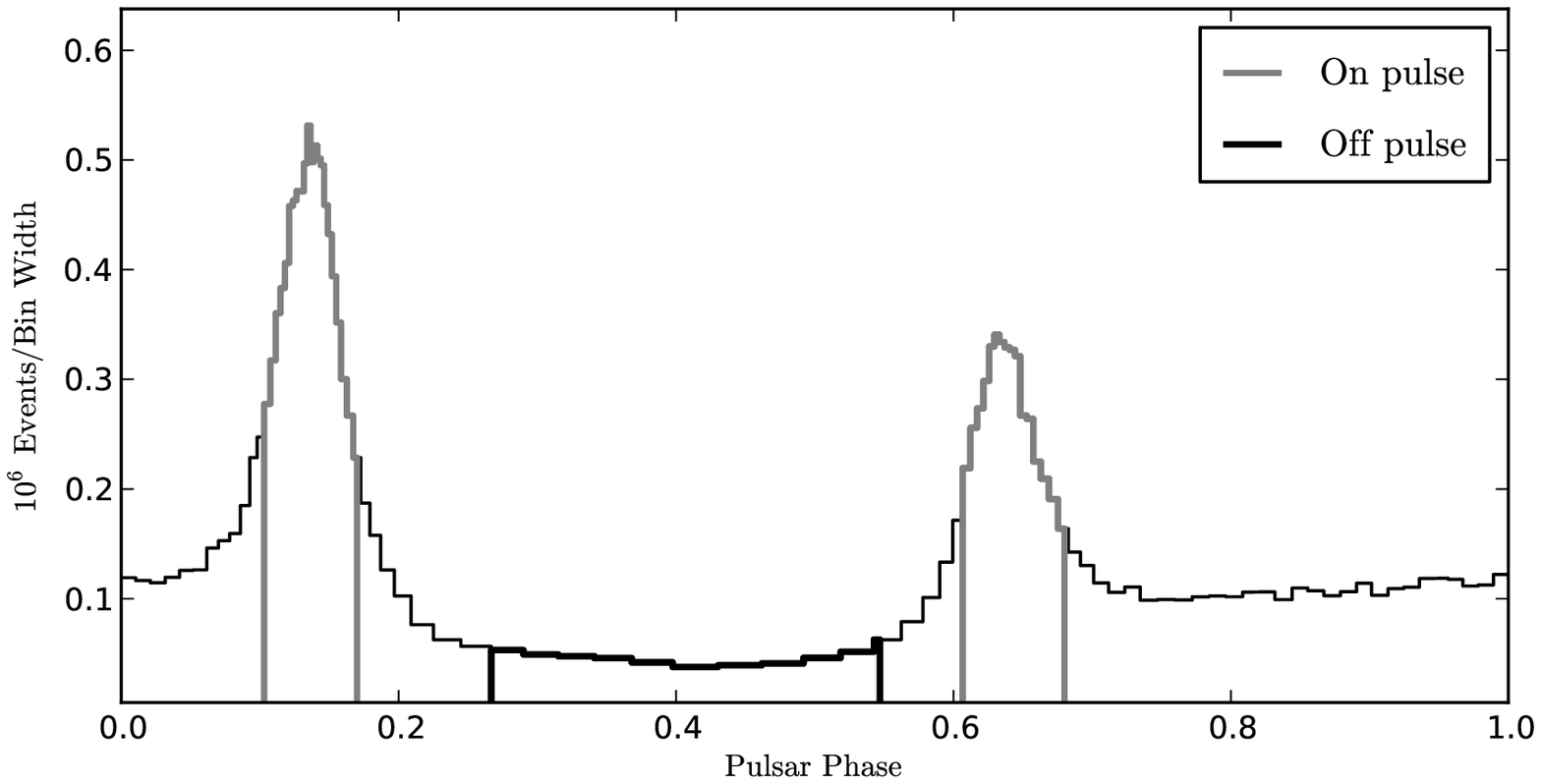}
\caption{Light curve for all $\gamma$ rays above 100 MeV within
  $4^{\circ}$ of the Vela pulsar (above) and Geminga pulsar (below) in 100 fixed-count phase bins. The
  on- and off-pulse phase selections are identified in grey and black,
  respectively.}
\label{FIGURE::gemphase}
\end{figure}

We verified the Monte Carlo (\irf{P6\_V3}) and on-orbit
(\irf{P6\_V11}) PSF models for the \textit{Diffuse} event class with the angular distributions
of $\gamma$ rays from the pulsars Geminga (PSR J0633+1746) and Vela (PSR
J0835$-$4510). The $\gamma$-ray sample was
divided into four logarithmic bins per decade in energy and also
separated into front and back
conversion types.
For each energy bin and conversion type, on- and off-pulse angular
distributions were created from the pulsar data sample by selecting
events with pulse-phase ranges given in Table
\ref{TABLE::PULSAR_PHASE}. Pulse phases for the Vela pulsar were determined by using
radio ephemerides derived from timing
with the Parkes telescope \citep{REF:2010.Weltevrede} and the pulse
phases for the Geminga pulsar were determined from \textit{Fermi} data
\citep{REF:2010.Geminga}.  The pulse phases were applied to LAT data
with the TEMPO2 application
\citep{REF:2006.Hobbs}\footnote{\textit{Fermi} pulsar ephemerides may
  be found at http://fermi.gsfc.nasa.gov/ssc/data/access/lat/ephems/}.
Light curves for Vela and Geminga can be seen in Figure
\ref{FIGURE::gemphase}.  The background in
the on-pulse distributions was estimated by scaling the off-pulse
distributions by the ratio of the widths of the on- and off-pulse
phase intervals.  The angular distribution of $\gamma$ rays from a
point source was then inferred as the differences by angular bin of
the on- and scaled off-pulse counts distributions.  This technique
should provide a perfect subtraction of any unpulsed sources of
$\gamma$-ray emission such as would be associated with a pulsar wind
nebula (PWN) or the Galactic diffuse emission.  The Vela-X PWN, a
spatially extended source that is offset from the Vela Pulsar by $\sim
1^\circ$, is approximately 500 times fainter than the Vela pulsar at 1
GeV \citep{REF:2010.VelaPWN}.  No evidence for a PWN has been found
associated with the Geminga pulsar \citep{2011ApJ...726...35A}.

Above 10 GeV where the statistics in the pulsar data set are limited,
a comparison was made with the same high-latitude AGN sample that was
used to derive the on-orbit PSF model.  For the stacked AGN sample,
the background was estimated by assuming an isotropic intensity
determined by the events in the annulus of angular radius range
1.5$-$3.0 deg centered on the stacked 1FGL coordinates of the blazars.
The 68\% and 95\% containment radii for the events in each energy bin
were measured from the cumulative distribution of the excess.  The
statistical errors on the containment radii derived from both pulsars
and stacked AGN were estimated from the dispersions of containment
radii determined from a large sample of Monte Carlo realizations for
the signal and background distributions.

\begin{deluxetable}{lll}
  \tablecaption{On- and off-pulse phase selection for Vela and Geminga
    pulsars. \label{TABLE::PULSAR_PHASE}}
\tablewidth{0pt}
\tablehead{\colhead{Source} & \colhead{On-pulse} & \colhead{Off-pulse} }
\startdata
Vela (PSR J0835$-$4510)& 0.1--0.15, 0.5--0.6 & 0.7--0.1 \\
Geminga (PSR J0633+1746) & 0.1--0.17, 0.6--0.68 & 0.25--0.55
\enddata
\end{deluxetable}

\begin{deluxetable}{llllll}
\tabletypesize{\scriptsize}

\tablecaption{68\% containment radii (degrees) of the LAT PSF for front
  events in the \textit{Diffuse} class as a function of energy bin
  inferred from different calibration data sets: Vela, Geminga, and
  the AGN calibration sample.  The containment radius of a source with
  the observing profile of Vela and a power-law energy distribution
  with a photon index $\Gamma = 2$ calculated with the \irf{P6\_V3}
  and \irf{P6\_V11} PSFs is shown for
  comparison. \label{TABLE::PULSAR_CONTAINMENT_FRONT}}

\tablewidth{0pt}
\tablehead{Energy Bin &\multicolumn{5}{c}{}\\
  \colhead{[log$_{10}$(E/MeV)]} & Vela & Geminga &AGN&P6\_V3 & P6\_V11}
\startdata
2.00 --  2.25& $2.62 \pm 0.06$   & $2.2 \pm 0.2$      &  \nodata              &          2.77         &     2.99             \\
2.25 --  2.50& $1.94 \pm 0.02$   & $2.02 \pm 0.09$    &  \nodata              &          1.88         &     1.96             \\
2.50 --  2.75& $1.24 \pm 0.01$   & $1.25 \pm 0.03$    &  \nodata              &          1.21         &     1.23             \\
2.75 --  3.00& $0.771 \pm 0.008$ & $0.78 \pm 0.01$    &  \nodata              &          0.754        &     0.763             \\
3.00 --  3.25& $0.48 \pm 0.005$  & $0.483 \pm 0.008$  &  \nodata              &         0.466         &     0.481             \\
3.25 --  3.50& $0.313 \pm 0.004$ & $0.301 \pm 0.005$  &  \nodata              &         0.300         &     0.309             \\
3.50 --  3.75& $0.205 \pm 0.004$ & $0.212 \pm 0.006$  &  $0.188 \pm 0.005$    &         0.201         &     0.209             \\
3.75 --  4.00& $0.173 \pm 0.009$ & $0.20 \pm 0.01$    &  $0.168 \pm 0.006$    &         0.139         &     0.154             \\
4.00 --  4.25& $0.15 \pm 0.01$   & $0.13 \pm 0.02$    &  $0.137 \pm 0.008$    &         0.0984        &     0.128             \\
4.25 --  4.50& $0.11 \pm 0.02$   & $0.19 \pm 0.07$    &  $0.113 \pm 0.007$    &         0.0723        &     0.116             \\
4.50 --  4.75& \nodata           & \nodata            &  $0.088 \pm 0.009$    &         0.0576        &     0.112             \\
4.75 --  5.00& \nodata           & \nodata            &  $0.08 \pm 0.01$      &         0.0516        &     0.110             \\
\enddata
\end{deluxetable}

\begin{deluxetable}{llllll}
\tabletypesize{\scriptsize}

\tablecaption{As in Table \ref{TABLE::PULSAR_CONTAINMENT_FRONT}, but for back events. \label{TABLE::PULSAR_CONTAINMENT_BACK}}

\tablewidth{0pt}
\tablehead{Energy Bin &\multicolumn{5}{c}{}\\
\colhead{[log$_{10}$(E/MeV)]} & Vela & Geminga &AGN&P6\_V3 & P6\_V11}
\startdata
2.00 --  2.25& $4.7 \pm 0.1$      & $5.0 \pm 0.5$    & \nodata             & 4.74     &  5.04        \\
2.25 --  2.50& $3.29 \pm 0.04$    & $3.2 \pm 0.1$    & \nodata             & 3.21     &  3.38        \\
2.50 --  2.75& $2.12 \pm 0.02$    & $2.1 \pm 0.05$   & \nodata             & 2.09     &  2.18        \\
2.75 --  3.00& $1.35 \pm 0.01$    & $1.41 \pm 0.03$  & \nodata             & 1.31     &  1.40        \\
3.00 --  3.25& $0.88 \pm 0.01$    & $0.89 \pm 0.01$  & \nodata             & 0.822    &  0.911        \\
3.25 --  3.50& $0.59 \pm 0.01$    & $0.6 \pm 0.01$   & \nodata             & 0.525    &  0.613        \\
3.50 --  3.75& $0.412 \pm 0.009$  & $0.44 \pm 0.01$  & $0.40 \pm 0.02$     & 0.347    &  0.440        \\
3.75 --  4.00& $0.37 \pm 0.02$    & $0.34 \pm 0.02$  & $0.36 \pm 0.02$     & 0.240    &  0.344        \\
4.00 --  4.25& $0.37 \pm 0.05$    & $0.29 \pm 0.04$  & $0.30 \pm 0.02$     & 0.175    &  0.299        \\
4.25 --  4.50& $0.3 \pm 0.2$      & $0.5 \pm 0.3$    & $0.19 \pm 0.02$     & 0.136    &  0.278        \\
4.50 --  4.75& \nodata            & \nodata          & $0.19 \pm 0.03$     & 0.113    &  0.269        \\
4.75 --  5.00& \nodata            & \nodata          & $0.23 \pm 0.05$     & 0.101    &  0.265        \\
\enddata
\end{deluxetable}

Tables \ref{TABLE::PULSAR_CONTAINMENT_FRONT} and
\ref{TABLE::PULSAR_CONTAINMENT_BACK} give the 68\% containment radii
for front and back events estimated from Geminga and Vela below 31.6
GeV and the AGN calibration data set above 3.16 GeV.  For comparison
the exposure- and spectrally-weighted PSF model prediction for the
\irf{P6\_V3} and \irf{P6\_V11} PSFs are shown for a source with the
observing profile of Vela and a power-law energy distribution ($dN/dE
\propto E^{-\Gamma}$) with a photon index of $\Gamma = 2$.  Although
the effective PSF depends on the observing profile, for observations
that span a time period many times greater than the orbital precession
period of the LAT (53.4 days), which is the case for this analysis,
the effective PSF model has only a weak dependence on the source
location on the sky and is primarily a function of declination.  In
the energy range 100 MeV -- 100 GeV, the largest difference between
the 68\% containment radii calculated from the \irf{P6\_V3} PSF using
the observing profiles of Vela and the stacked AGN sample is 2\%, and
we therefore assume that the differences in the observing profiles of
the calibration data sets can be ignored for the purposes of these
comparisons.  We find that the AGN and pulsar data sets analyzed here
give consistent estimates of the PSF size as a function of energy.
The agreement between the containment radii inferred from Geminga and
Vela validates the approach of using the off-pulse events to define
the background.

\begin{figure}[htb]
\centering
\includegraphics[width=0.70\textwidth]{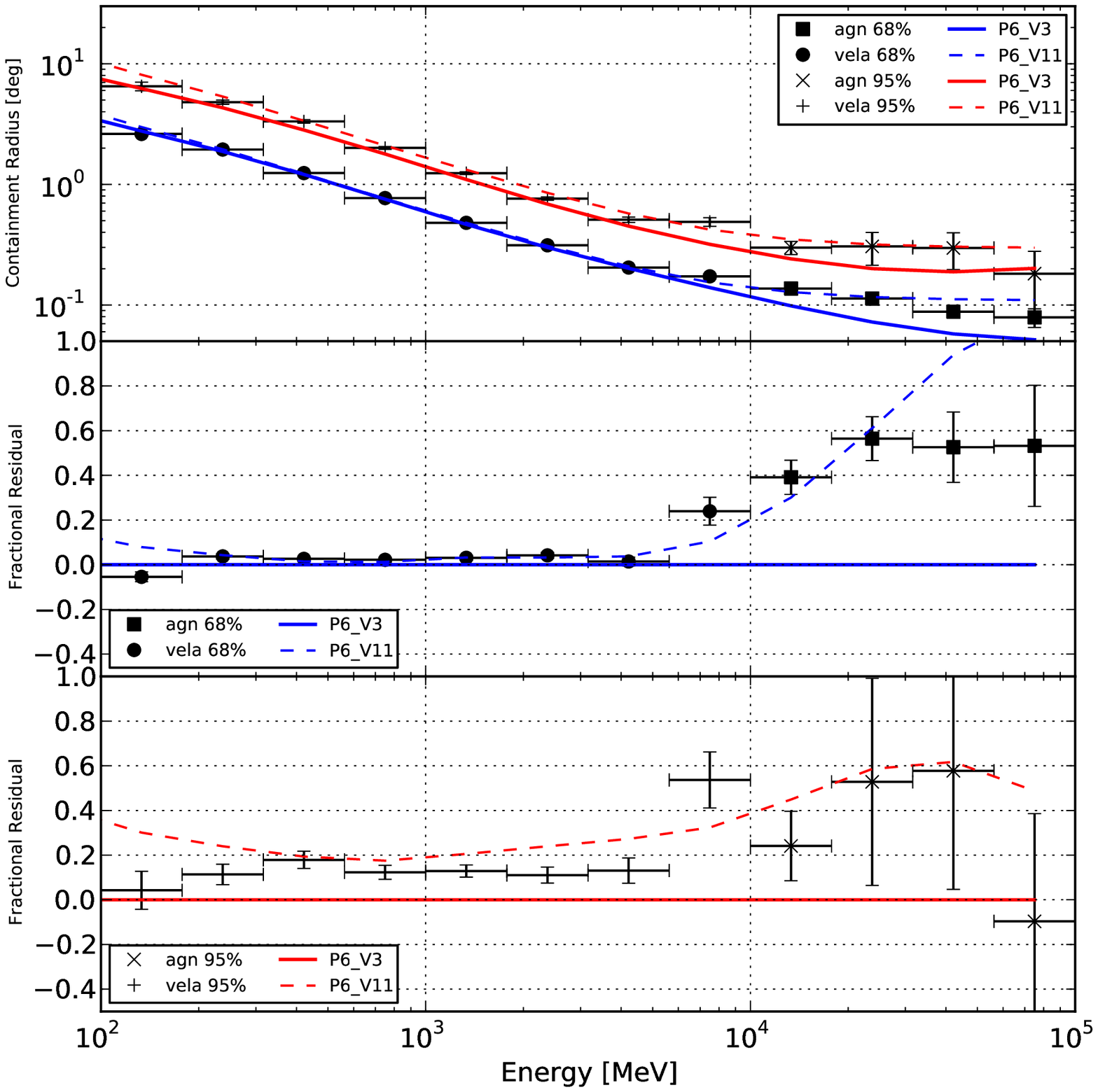}
\caption{Containment radii for front-converting \textit{Diffuse}-class
  events as determined from the angular distributions of Vela and the
  stacked AGN sample.  Blue and red curves show the 68\% and 95\%
  containment radii, respectively, given by the model predictions for
  the \irf{P6\_V3} (solid curve) and \irf{P6\_V11} (dashed curve)
  PSFs.  PSF model predictions are shown for an observing profile
  corresponding to the Vela pulsar and a power-law energy distribution
  with photon index $\Gamma = 2$. The middle and lower panels show the
  fractional residuals of the 68\% and 95\% containment radii of Vela,
  the stacked AGN sample, and the \irf{P6\_V11} PSF relative to the
  \irf{P6\_V3} PSF.}
\label{FIGURE::PULSAR_CONTAINMENT_front}
\end{figure}

\begin{figure}[htb]
\centering
\includegraphics[width=0.70\textwidth]{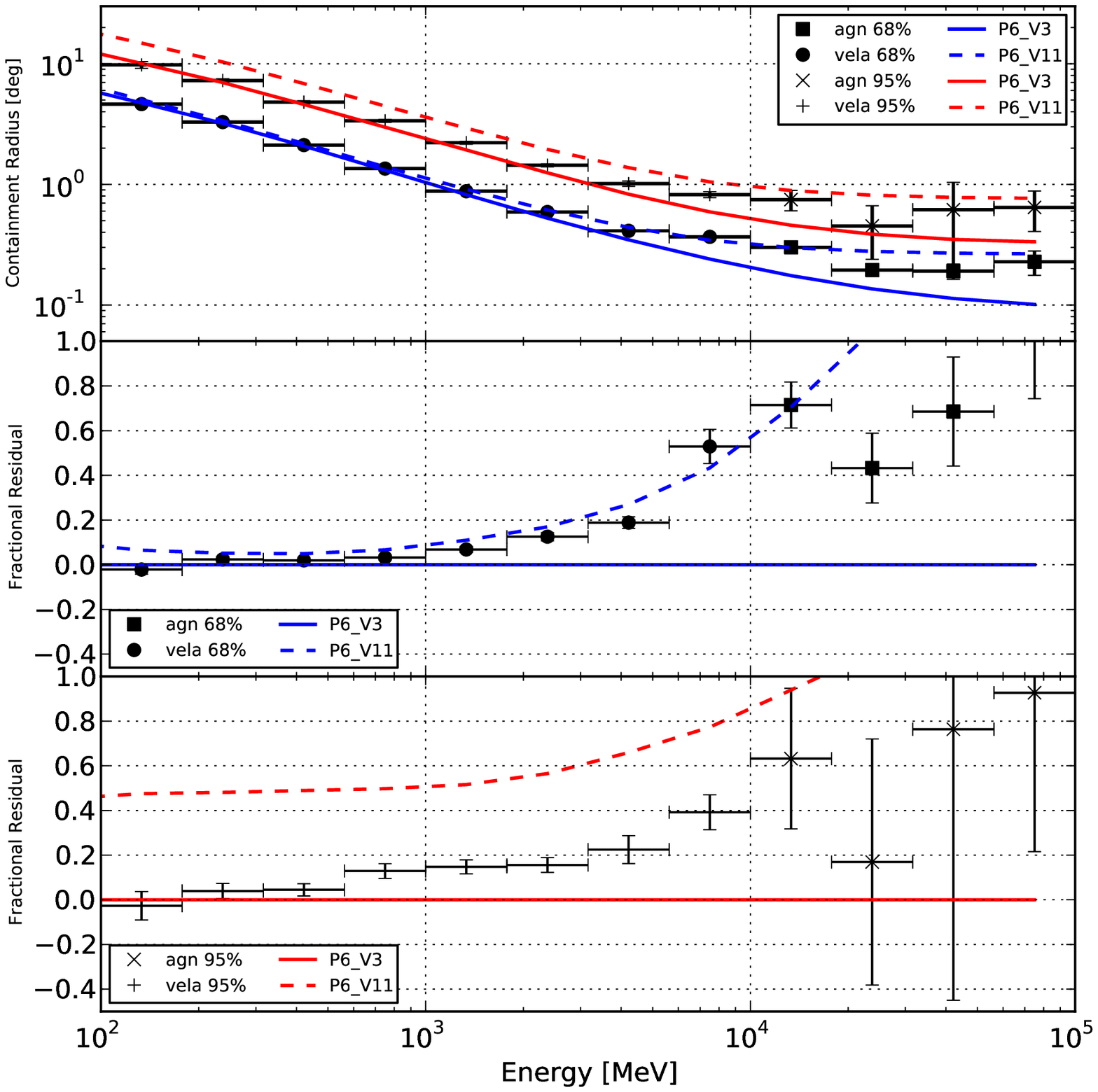}
\caption{Same as for Figure \ref{FIGURE::PULSAR_CONTAINMENT_front},
  but for back-converting events. }
\label{FIGURE::PULSAR_CONTAINMENT_back}
\end{figure}

Figures \ref{FIGURE::PULSAR_CONTAINMENT_front} and
\ref{FIGURE::PULSAR_CONTAINMENT_back} compare the PSF containment
radii inferred from Vela and the AGN calibration data sets
as a function of energy with the containment
radii given by the \irf{P6\_V3} and \irf{P6\_V11} PSFs for
front and back events, respectively.  The residuals
of both PSFs in the 68\% containment radius are less than 10\% below 3
GeV for both event classes.  Above 3 GeV the MC PSF model
(\irf{P6\_V3}) begins to significantly underpredict the size of the
68\% containment radius of both front and back
events.  The on-orbit PSF model (\irf{P6\_V11}) provides an improved
representation of the 68\% containment radius at high energies with
residuals less than 20\% but overpredicts the 95\% containment radius
for back events.  We attribute the \irf{P6\_V11}
back residual to using the single King function
parameterization, which can overestimate the 95\% containment radius
of the PSF (see for example Figure \ref{FIGURE::PSFDIFF}).

\subsection{Systematic Uncertainties in Modeling the LAT PSF}

Although the reason for the discrepancy between the on-orbit
(\irf{P6\_V11}) and MC (\irf{P6\_V3}) PSFs at high energies is not
fully understood, we argue that it is not due to intrinsic broadening
of the $\gamma$-ray distributions around the AGN sample that was used
for calibration.  Above 3 GeV the discrepancy in the 68\% containment
is $0.1-0.2^\circ$ for back events but $< 0.1^\circ$ for front events
(see Tables \ref{TABLE::PULSAR_CONTAINMENT_FRONT} and
\ref{TABLE::PULSAR_CONTAINMENT_BACK}).  Given that the front PSF is
approximately two times narrower than the back PSF, this discrepancy
cannot be self-consistently modeled as an intrinsic spatial extension
convolved with the LAT PSF.  Furthermore, in the intermediate energy
range (3 GeV -- 30 GeV) where the PSF can be independently measured
using both pulsars and AGNs, the PSF containments inferred from
pulsars are found to be consistent with those inferred from AGNs (see
Tables \ref{TABLE::PULSAR_CONTAINMENT_FRONT} and
\ref{TABLE::PULSAR_CONTAINMENT_BACK}).  We therefore conclude that the
the majority of the PSF discrepancy can be attributed to systematic
uncertainty in modeling the LAT.


We considered the residuals in the boresight alignment as a potential
source of systematic uncertainty.
The boresight alignment of the LAT is the orientation of its
coordinate system with respect to the spacecraft coordinate system.
The spacecraft determines its orientation with a pair of star trackers
with an accuracy of a few arcseconds.
The boresight alignment of the LAT is determined from analysis of
bright point sources accumulated over several months
\citep{REF:2009.OnOrbitCalib}.  We determined the boresight alignment 
early on in the mission, and the magnitude was measured to be $0.15^{\circ}$
and has been monitored on weekly and monthly bases for the entire
time range of the data considered here.  The fluctuations are less
than $0.005^{\circ}$ in a month. Therefore we rule out variations in the 
boresight alignment as contributing to the broadening of the on-orbit PSF 
relative to the MC PSF.


In the LAT event reconstruction software \citep{REF:2009.LATPaper},
positional and directional information from the Calorimeter (CAL)
detector system is used to seed the pattern recognition analysis that
is applied to track candidates recorded by the Tracker (TKR) detector
system.  Furthermore, the vector from the centroid of the energy deposition 
in the CAL to the estimated $\gamma$-ray conversion point is considered 
as an additional constraint on the direction of the incoming $\gamma$ ray above $\sim 1$~GeV. 
The calculation of position
information from the CAL relies on accurate maps of the scintillation
response of the CAL crystals
\citep{REF:2009.LATPaper,2010SPIE.7732E..16G}.  The response maps used
to produce the Pass6 data release were derived prior to launch using
cosmic ray muons.


We identified the crystal response maps as a possible source of the
discrepancy between the observed and simulated PSF at high energies,
either because of a time dependence in the crystal response or because
of an inaccuracy of the maps in representing the actual spatial
dependence of the crystal response.  To evaluate whether the PSF
discrepancy was changing with time, we determined the 68\% and 95\%
containment radii for each of the energy bins in Section
\ref{SECTION::ONORBITANALYSIS} in six 5-month intervals.  We detected
no significant changes in the containment radii in the energy range
1$-$32 GeV for either conversion type.  Over the same time interval,
however, on-orbit radiation damage to the scintillating crystals
caused a typical decrease in scintillation light attenuation length of
about 3\%, which corresponds to an average position bias of ~3 mm near
the ends of crystals and up to 10 mm bias in the most sensitive
crystals.  Because the PSF did not show a detectable change with time,
we conclude that time dependence in the crystal response is not the
dominant source of the PSF discrepancy.


To test whether inaccuracy in the response maps could be the cause, we
reanalyzed the sea-level and on-orbit cosmic ray calibration data to
derive maps that more closely describe the response near the end of
each crystal.  Using the revised response maps, we repeated the event
reconstruction for a test data set consisting of events from five AGN
from the calibration sample at high latitude toward directions with
low intensities of Galactic diffuse emission.  For $\gamma$ rays with
energy greater than 5 GeV, we found that the mean angular separation
from the source position in this event sample drops from
$0.133^{\circ} \pm 0.004^{\circ} $ to $0.114^{\circ} \pm
0.004^{\circ}$.  By using the improved calibration, we recovered
$\sim$ 70\% of the resolution loss relative to the Monte Carlo
expectation.  We conclude, therefore, that inaccuracy in the crystal
response maps used for the Pass6 event reconstruction indeed is the
source of much of the PSF discrepancy.  More detailed analysis and
diagnosis is in progress.

\section{Pair-Halo Analysis}
As discussed in Section 1, the IGMF broadens the angular extent of
$\gamma$-ray emission from AGN into pair halos through the deflection
of electromagnetic cascades.  In contrast, the pulsed $\gamma$-ray
emission from any pulsar appears as a true point source to the LAT and
the pulsar emission can be effectively separated from the background
of any surrounding nebula and of the diffuse interstellar and
extragalactic emission through the phase-based background subtraction
technique described in section \ref{SECTION::ONORBITVALID}.  In the
following sections, we place limits on the angular extension of AGN
emission relative to pulsar emission and present an analysis that
evaluates the significance of two extended angular profiles for BL Lac
blazar populations and TeV sources, using pulsars as calibration
sources.

\subsection{Maximum Likelihood Analysis in Angular Bins}

To test for the presence of pair halos around AGN, we use a joint likelihood for the angular distributions of $\gamma$ rays around AGN and pulsars. 
The events are binned into three logarithmic energy intervals from 1$-$31.6 GeV.
Additionally, the sample is binned in angular offset from their presumed source such that there are an equal number of counts in each of twelve angular bins for the on-pulse counts from the Vela pulsar.
Since the Vela pulsar emission is an order of magnitude brighter than the background intensity in the on-pulse phase, this choice of binning ensures that the integrated efficiency in each bin and thus the point-source statistics are roughly the same for all sources in each angular bin.
The front converting events have lower rates of residual cosmic rays
 and better angular resolution than the back
events, and therefore we limit the analysis to these events. 

We used a non-parametric representation of the PSF given by the fraction of events ($m_i$) in each of the 12 angular bins, providing a 
more direct comparison between the pulsar and AGN angular
distributions by removing any dependence of the analysis on the choice
of PSF parameterization.
The model for the angular distribution of events for the on-pulse pulsar emission, ${\nu}^{on}_i$, is expressed as
\begin{equation}
{\nu}^{on}_i = N^{psr}m_i + \alpha~\nu^{off}_i\;,
\label{eq:onpulse}
\end{equation}
where $N^{psr}$ is the number of events attributed to the pulsar in the on-pulse phase, $\nu^{off}_i$ is the model for the number of off-pulse events in angular bin $i$, $m_i$ is the PSF weight in angular bin $i$, and $\alpha$ is the ratio of the width of the on and off-pulse phase selections.
We chose Vela and Geminga as calibration sources for this analysis, as these pulsars have have the largest number of source $\gamma$ rays above 100 MeV and weak or undetected associated nebular emission. 
The on- and off-pulse data samples were defined using the phase ranges from Table \ref{TABLE::PULSAR_PHASE} and the angular bin ranges, counts, and models are shown in Tables \ref{TABLE::PULSARANG} $-$ \ref{TABLE::PULSARANG2}.

The model of the angular distribution of $\gamma$ rays from AGN is the sum of three components: point-source emission, a uniform background, and extended (halo) emission. It is given by
\begin{equation}
\nu^{agn}_i =N^{agn}m_i + N^{iso}b_i + N^{halo}h_i^*(\theta_0)\;,
\label{eq:agncomp}
\end{equation}
where $N^{agn}$ is the total number of events attributed to the AGN, $N^{iso}$ and $b_i$ are the total number and fraction of events in angular bin $i$ for the isotropic model. $N^{halo}$ and $h_i^*(\theta_0)$ are the total number and fraction of events in angular bin $i$ for the halo model, $h_i$, convolved with the PSF. 
The isotropic fractions, $b_i$, were calculated from the fraction of solid angle in the ROI contained in angular bin $i$. 
For the halo models tested in this work, $h_i$ has a single parameter,  $\theta_0$, corresponding to a characteristic halo size. 
We convolved the halo model with a single King function that was fit to the PSF weights in the null halo case. 

Given the observations $\vec{n}^{on}, \vec{n}^{off}$, and $\vec{n}^{agn}$ corresponding to the on- and off-pulse pulsar and AGN counts in each of the 12 angular bins, the joint likelihood for the stacked pulsars and AGN is
\begin{equation}
\begin{split}
\textrm{log}L(\vec{m},\vec{b}, \alpha, N^{psr}, \vec{\nu}^{off}, N^{agn}, & N^{iso}, N^{halo},\theta_0 \mid \vec{n}^{on}, \vec{n}^{off}, \vec{n}^{agn} ) = \\
& \sum^{angular~bins}_{i}  \textrm{log}L_P( ~{\nu}^{on}_i(m_i,N^{psr},\nu^{off}_i,\alpha)~\mid~n^{on}_i )~+ \\
& \sum^{angular~bins}_{i}   \textrm{log}L_P( ~\nu^{off}_i~ \mid~n^{off}_i)~+\\ 
& \sum^{angular~bins}_{i}   \textrm{log}L_P(~\nu^{agn}_i(m_i,b_i,N^{agn},N^{iso},N^{halo},\theta_0)~ \mid ~n^{agn}_i)\;,
\end{split}
\label{eq:binnedlike}
\end{equation}
where
\begin{equation}
\textrm{log}L_P(\nu \mid n) = n \log\nu - \nu - \log n!
\label{eq:poissonlike}
\end{equation}
is the log-likelihood for observing $n$ events given a model
amplitude $\nu$.  The maximum likelihood estimators of the model
parameters were evaluated by maximizing the joint likelihood with respect to
all model parameters given the data.

Various models for the angular profile of halo emission induced by the IGMF have been considered \citep{2009PhRvD..80b3010E,REF:2010.AndoKus,REF:2010.MAGIC}. Here we consider both Gaussian and Disk models; these can be expressed as
\begin{equation}
\frac{dN_{gauss}}{d\Omega} = h(\theta,\theta_0) = \frac{1}{\pi\theta_0^2} \exp \left( -\frac{\theta^2}{\theta_0^2}\right)
\label{eq:gausshalo}
\end{equation}
and 
\begin{equation}
\frac{dN_{disk}}{d\Omega} = h(\theta,\theta_0) \propto \begin{cases}
\frac{1}{\pi\theta_0^2} & \theta < \theta_0 \\
0 & \theta > \theta_0\;.
\end{cases}
\label{eq:diskhalo}
\end{equation}
where both equations are normalized with the small-angle approximation, $d\Omega=2\pi\theta d\theta$.
The Disk and Gaussian models were chosen to bracket the shape of the model tested by \citet{REF:2010.AndoKus}, with the Disk and Gaussian respresenting the limiting cases of a sharply peaked and broad distribution, respectively.
A test statistic $TS$ for the halo models as a function of $\theta_{0}$ was constructed by evaluating the difference between the maximum likelihood of the halo model ($L(N^{halo},\theta_{0})$) and the maximum likelihood of the null-hypothesis ($L(0,\theta_{0})$), 

\begin{equation}
TS_{halo}(\theta_{0}) = 2 (\log L(N^{halo},\theta_{0}) - \log L(0,\theta_{0}))
\label{eq:TS}
\end{equation}
where all parameters besides $\theta_0$ are left free. 
Given the constraint that $N^{halo}>0$ and the null case is on the boundary of the parameter space (i.e. $N^{halo}=0$), the significance of the pair halo component with characteristic extension $\theta_0$ is
$S = \sqrt{TS}\sigma$,
provided that the number of events associated with the pulsars and AGN, $N^{psr}$ and $N^{agn}$, is larger than $\sim 20$ \citep{1996ApJ...461..396M,REF:2002.Protassov}.

\subsection{Limits on the pair-halo emission of 1FGL BL Lac Sources}

In the \textit{Fermi} LAT First AGN Catalog catalog \citep[henceforth 1LAC,][]{REF:2010ApJ...715..429A}, 115 of the BL Lac-type AGN have measured redshifts. 
The sources were split into low and high redshift groups defined by $z<0.5$ and $z>0.5$, respectively, to test for a redshift-dependent size difference, e.g., \citet{REF:2010.AndoKus}. 
The number of low- and high-redshift sources is 94 and 21, respectively. 
For the AGN, we used our 2-year sample of \irf{P6\_V3} \textit{Diffuse}-class events, while the data for Vela and Geminga pulsars were further constrained by the time ranges of the available timing solutions, leaving 1.5 years of data for Vela and 1.4 for Geminga.

\begin{figure}[htb]
\centering

    \includegraphics[width=\twothirdscolfigwidth]{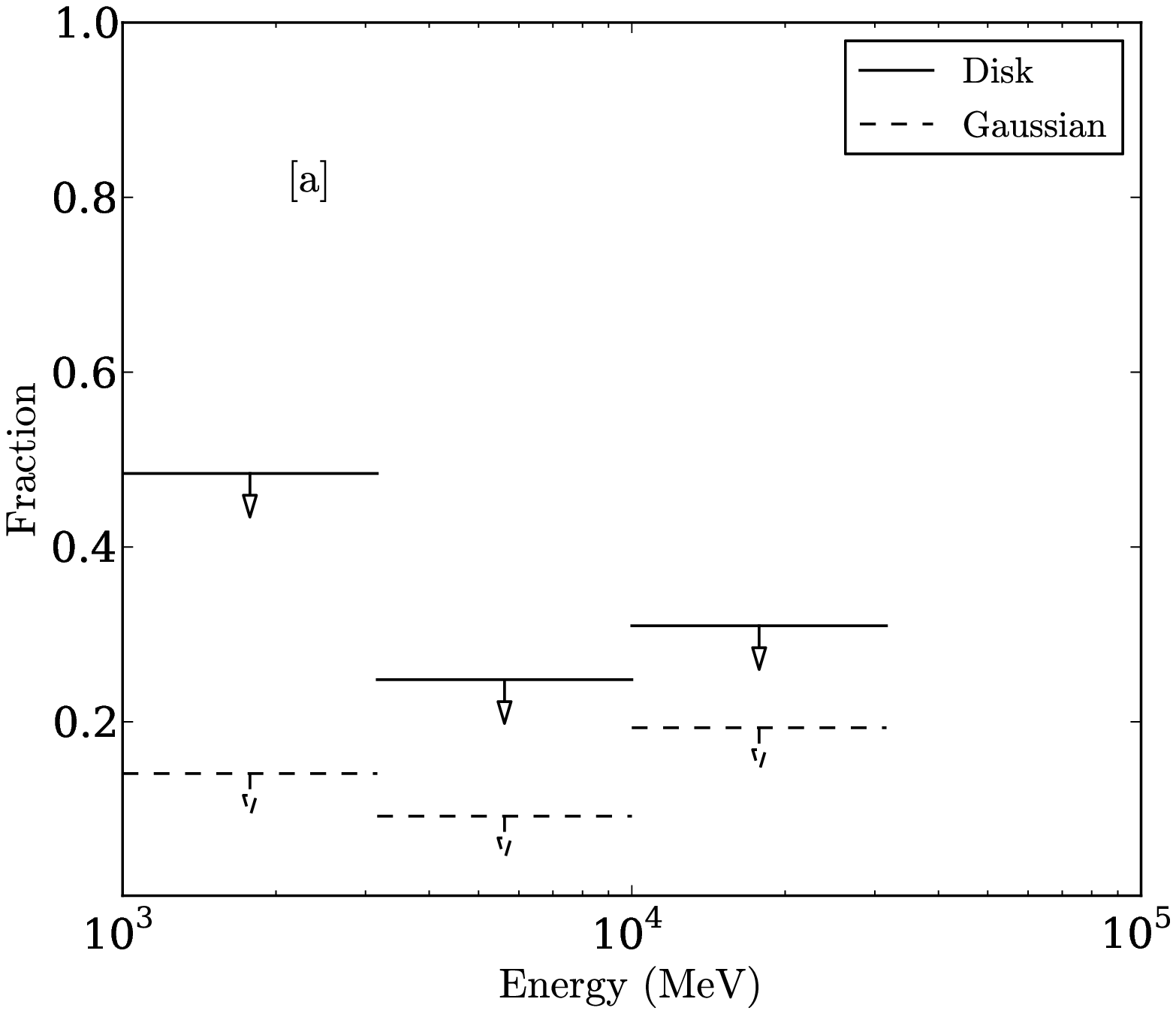}\hfill%
    \includegraphics[width=\twothirdscolfigwidth]{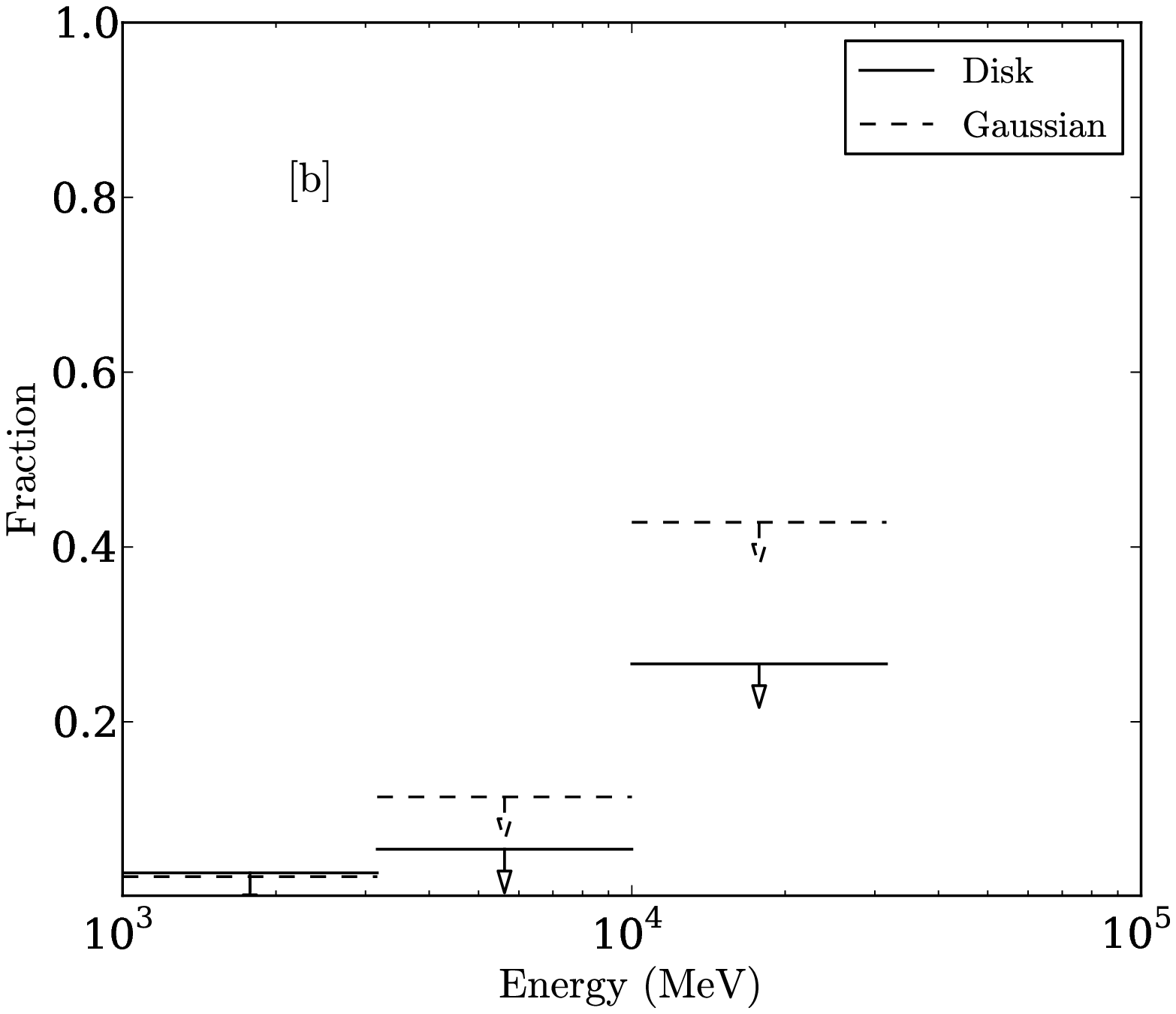}\hfill%
    \includegraphics[width=\twothirdscolfigwidth]{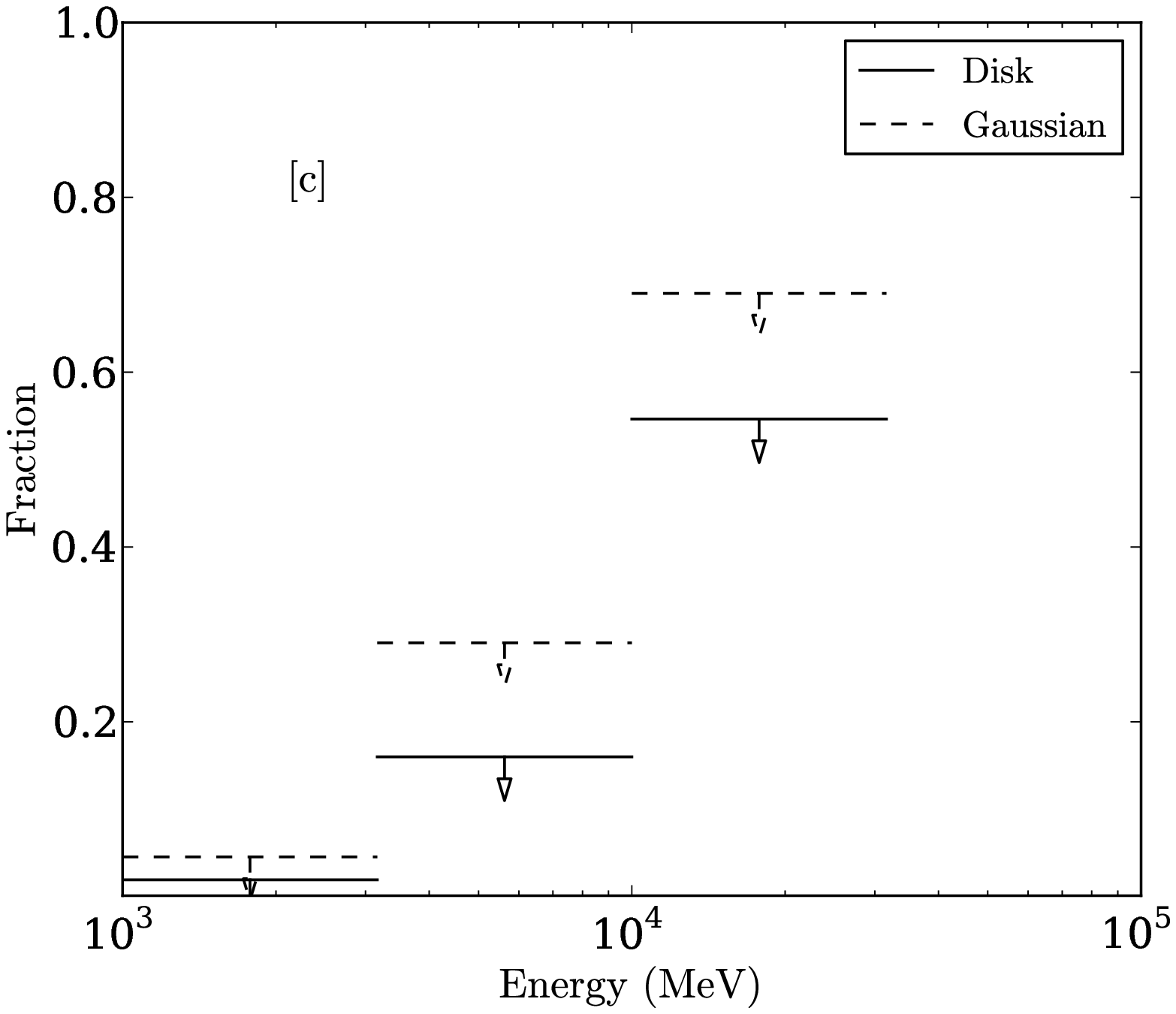} 
\caption{95\% upper limits on the fraction of halo model $\gamma$ rays from the low-redshift BL Lacs, assuming a 0.1$^\circ$ (\textbf{a}), 0.5$^\circ$ (\textbf{b}), and 1.0$^\circ$ (\textbf{c}) halo.
}
\label{FIGURE::agnlohalo}
\end{figure}

\begin{figure}[htb]
\centering
    \includegraphics[width=\twothirdscolfigwidth]{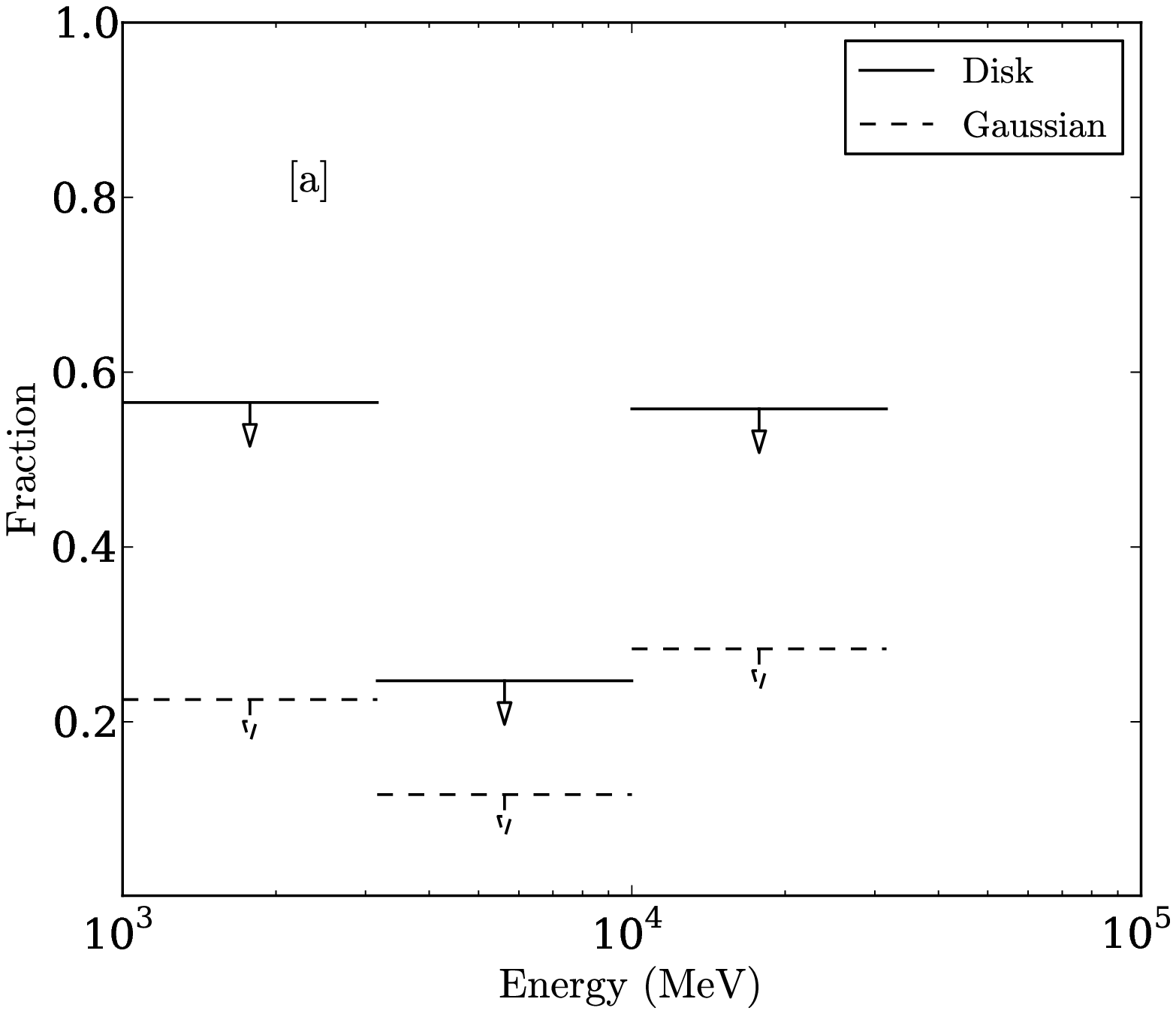}\hfill%
    \includegraphics[width=\twothirdscolfigwidth]{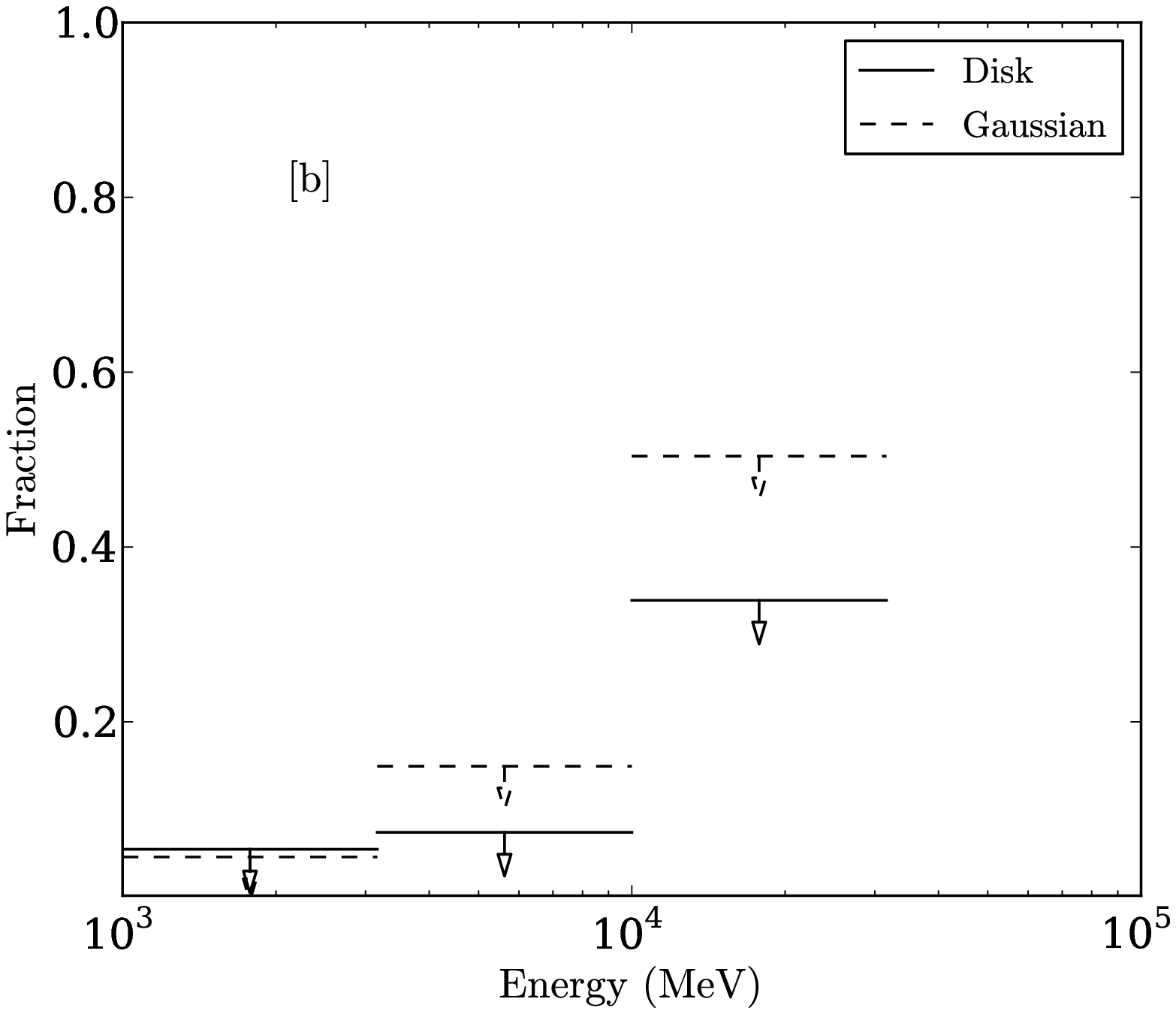}\hfill%
    \includegraphics[width=\twothirdscolfigwidth]{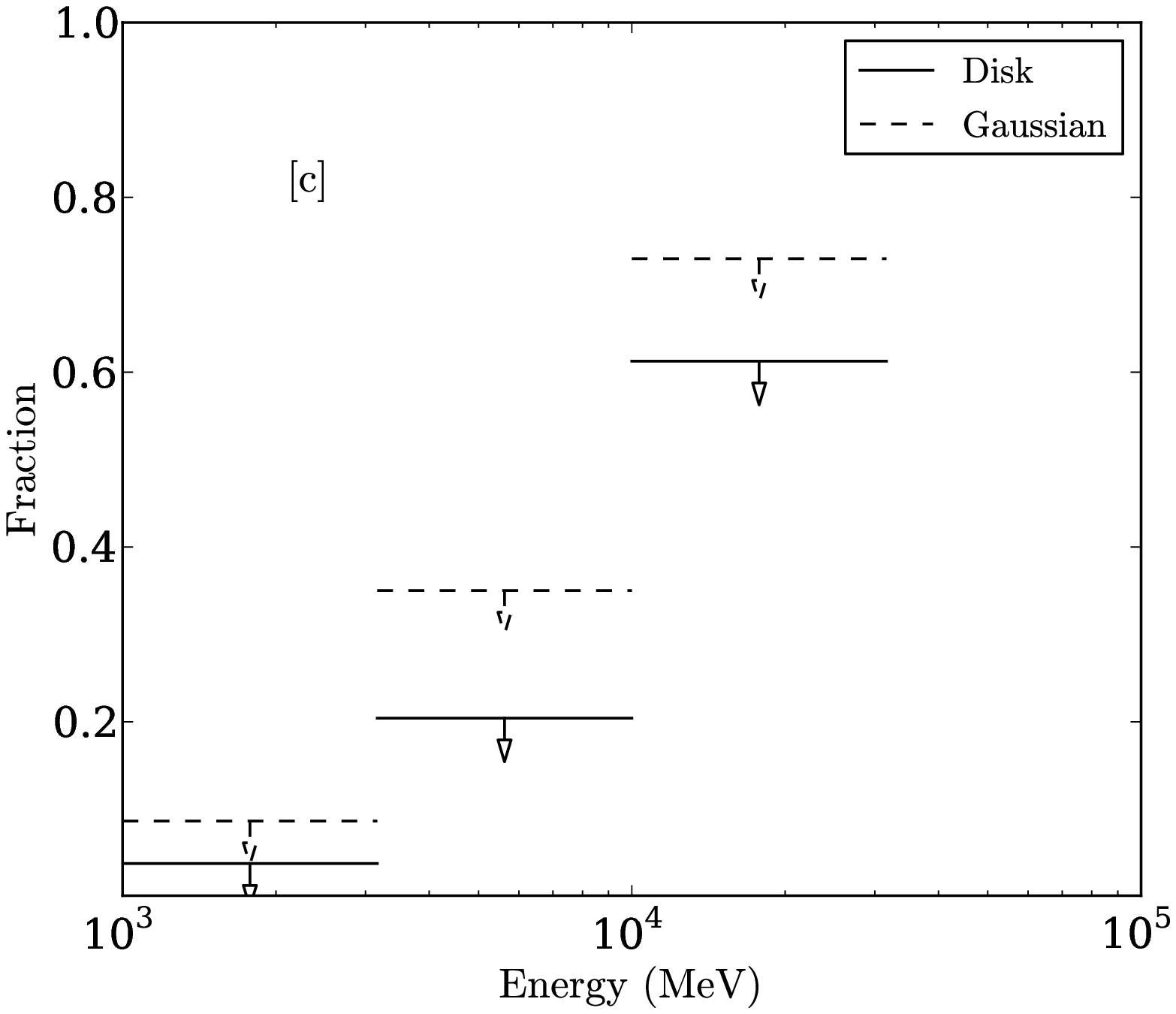} 
\caption{95\% upper limits on the fraction of halo model $\gamma$ rays from the high-redshift BL Lacs, assuming a 0.1$^\circ$ (\textbf{a}), 0.5$^\circ$ (\textbf{b}), and 1.0$^\circ$ (\textbf{c}) halo. 
}
\label{FIGURE::agnhihalo}
\end{figure}

As in Section 3, for the AGN we
included events in the energy range above 1 GeV to limit
the contamination from nearby bright sources.
The $\gamma$-ray data sets for the two redshift ranges were binned in
energy with two bins per logarithmic decade.  The significance of the
halo component was evaluated with the likelihood defined in Equation
\ref{eq:binnedlike}, and the Gaussian and Disk halo parameters
$\theta_0$ = 0.1, 0.5, and 1 degrees.  No TS larger than 0.1 ($S
\approx 0.3\sigma$) was obtained for any of the redshift sets or halo
parameters, so upper limits were derived for the fraction of
$\gamma$-rays from the stacked sample attributable to a halo component
($f_{halo}=N^{halo}/(N^{halo}+N^{agn}))$. This finding is in contrast
to the results of \citet{REF:2010.AndoKus}, who found $3.5~\sigma$
significance for $0.5-0.8^{\circ}$ extension ($f_{halo}=0.073$) in the
3 -- 10 GeV range for one year of LAT data for all 1FGL low-redshift
AGN ($z<0.5$) using front- and back-converting
events.  Over the same range of energy, a halo component of this
magnitude ($f_{halo}=0.073$) and angular size is excluded at the
$1.5~\sigma$ and $2.7~\sigma$ levels for the $0.5^{\circ}$ Gaussian
and Disk models, respectively.  The upper limits on $f_{halo}$ are
summarized in Tables \ref{TABLE::lzsum} and \ref{TABLE::hzsum} and
plotted in Figures \ref{FIGURE::agnlohalo} and
\ref{FIGURE::agnhihalo}.  For the smallest halo size, $\theta_0 =
0.1^{\circ}$, the primary background for the halo component is the
point-source $\gamma$ rays and the upper limits become less
constraining at low energies where the PSF is significantly broader
than the halo ($R_{68} \simeq 0.4^{\circ}$ at 1--3.16 GeV).
Sensitivity to the broader halo models, $\theta_0 = 0.5^{\circ}$ and
$1.0^{\circ}$, is limited by the isotropic background and thus the
upper limits for the broader Gaussian models are less constraining for
all energies.


\subsection{Limits on the pair-halo emission of TeV BL Lacs}

The TeV BL Lac-type AGN 1ES0229$+$200 ($z=0.140$) and 1ES0347$-$121
($z=0.188$) are predicted to have detectable emission in the LAT
energy range due to the suppression of the primary TeV $\gamma$ rays
from these sources by the EBL
\citep{2005ApJ...631..762W,REF:2010.Neronov2,REF:2010.Tavecchio,REF:2011.DermerHalo}.
The multi-TeV primary photons are converted into leptons that scatter
CMB photons to GeV energies, unless the IGMF is sufficiently strong to
deflect enough secondary pairs away from our line of sight
\citep{REF:2010.Neronov2}.  The blazar 1ES0229$+$200 provides the
strongest constraint on the IGMF due to its significant TeV emission
which extends to $\sim$ 10 TeV.  The inferred primary spectra from
synchrotron self-Compton emission are at or below the LAT sensitivity
for an observation of two years, leaving the secondary processes as
the primary detectable $\gamma$ rays from these sources
\citep{REF:2007A&A...473L..25A,REF:2011.DermerHalo}.  A detection of
1ES0229$+$200 in Fermi-LAT data was recently reported in
\citet{2012ApJ...747L..14V} using 39 months of data.
In the two year data set used for this analysis which includes
only front-converting events, we find no evidence for significant
point-like emission from this source.
We fit 1ES0229$+$200 and 1ES0347$-$121 with the PSF-convolved Disk and
Gaussian models with $\theta_0 = 0.1^{\circ}, 0.5^{\circ}, $ and
$1.0^{\circ}$ in four logarithmic energy bins between 1 and 100 GeV.
In the 32--100 GeV energy bin the number of $\gamma$ rays from pulsars
is insufficient to provide a template for the PSF, and we therefore
used low-redshift BL Lacs to define the PSF template in this energy
range.  The AGN angular bin ranges, counts, and model amplitudes are
shown in Table \ref{TABLE::AGNANG}.  Because 1ES0229$+$200 and
1ES0347$-$121 were not detected as a point source in the LAT energy
band, we tested only for the significance of the halo component by
setting $N^{agn}=0$ in Equation \ref{eq:binnedlike}.

\begin{figure}[htb]
\centering
    \includegraphics[width=\twothirdscolfigwidth]{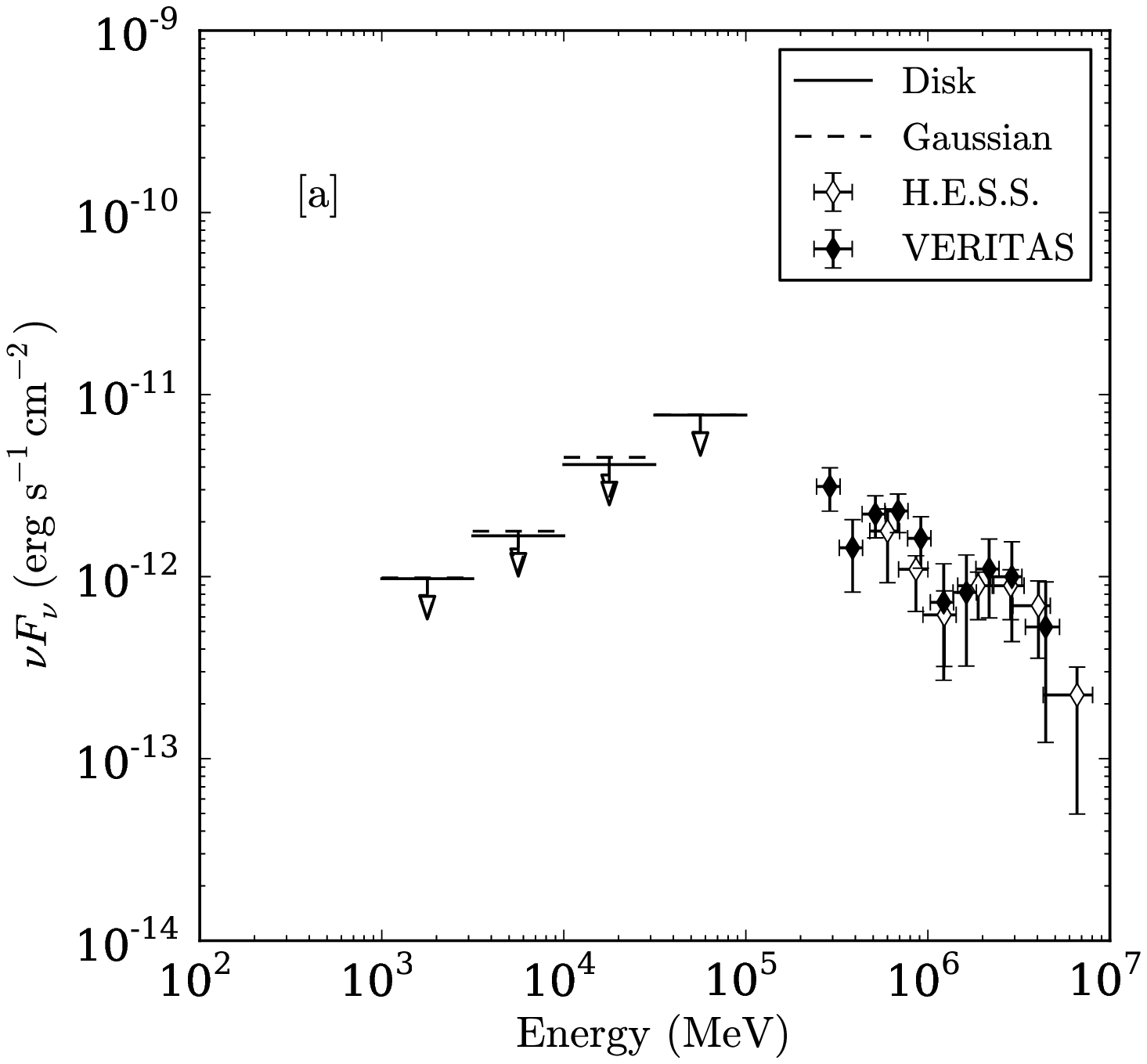}\hfill%
    \includegraphics[width=\twothirdscolfigwidth]{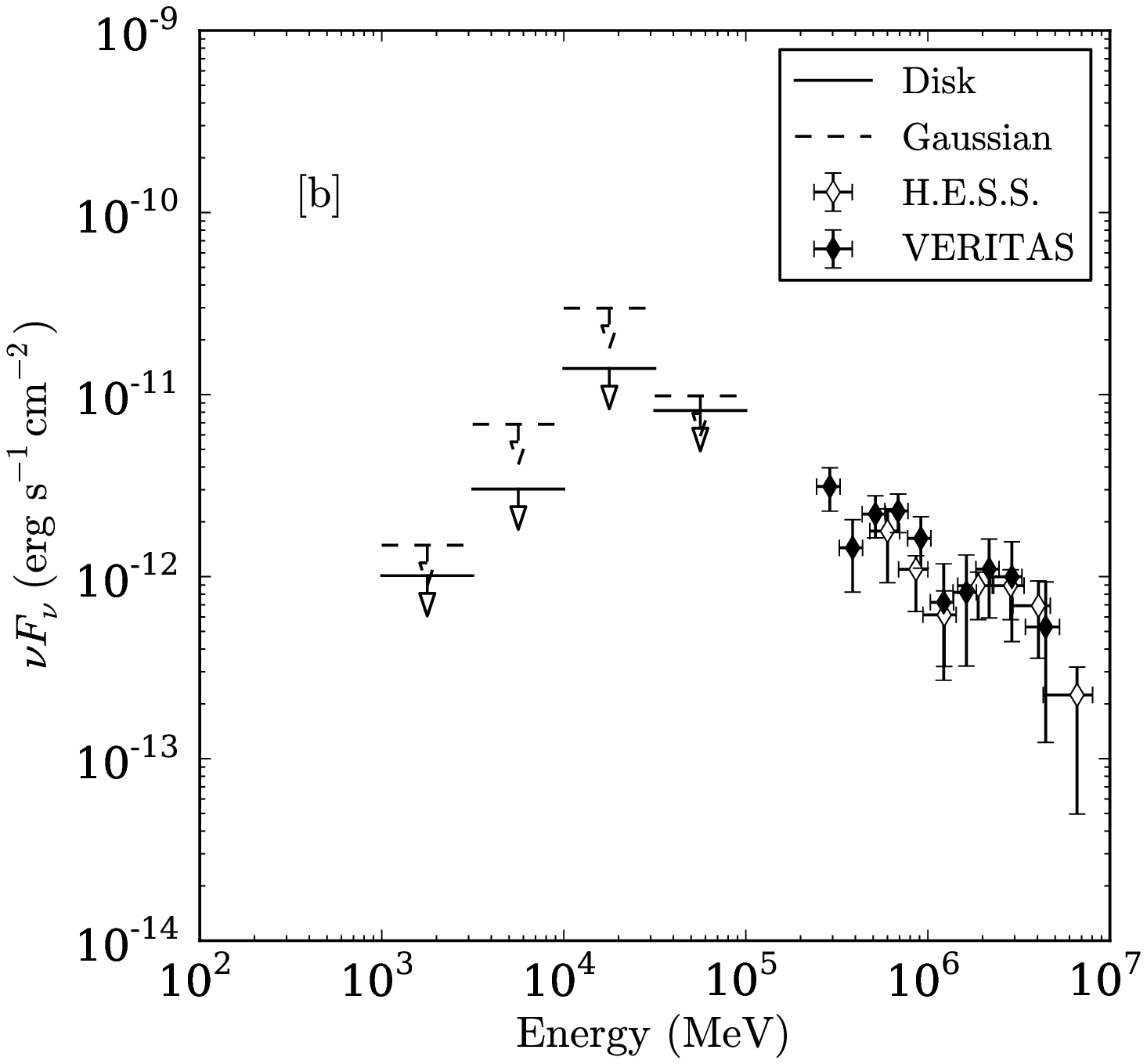}\hfill%
    \includegraphics[width=\twothirdscolfigwidth]{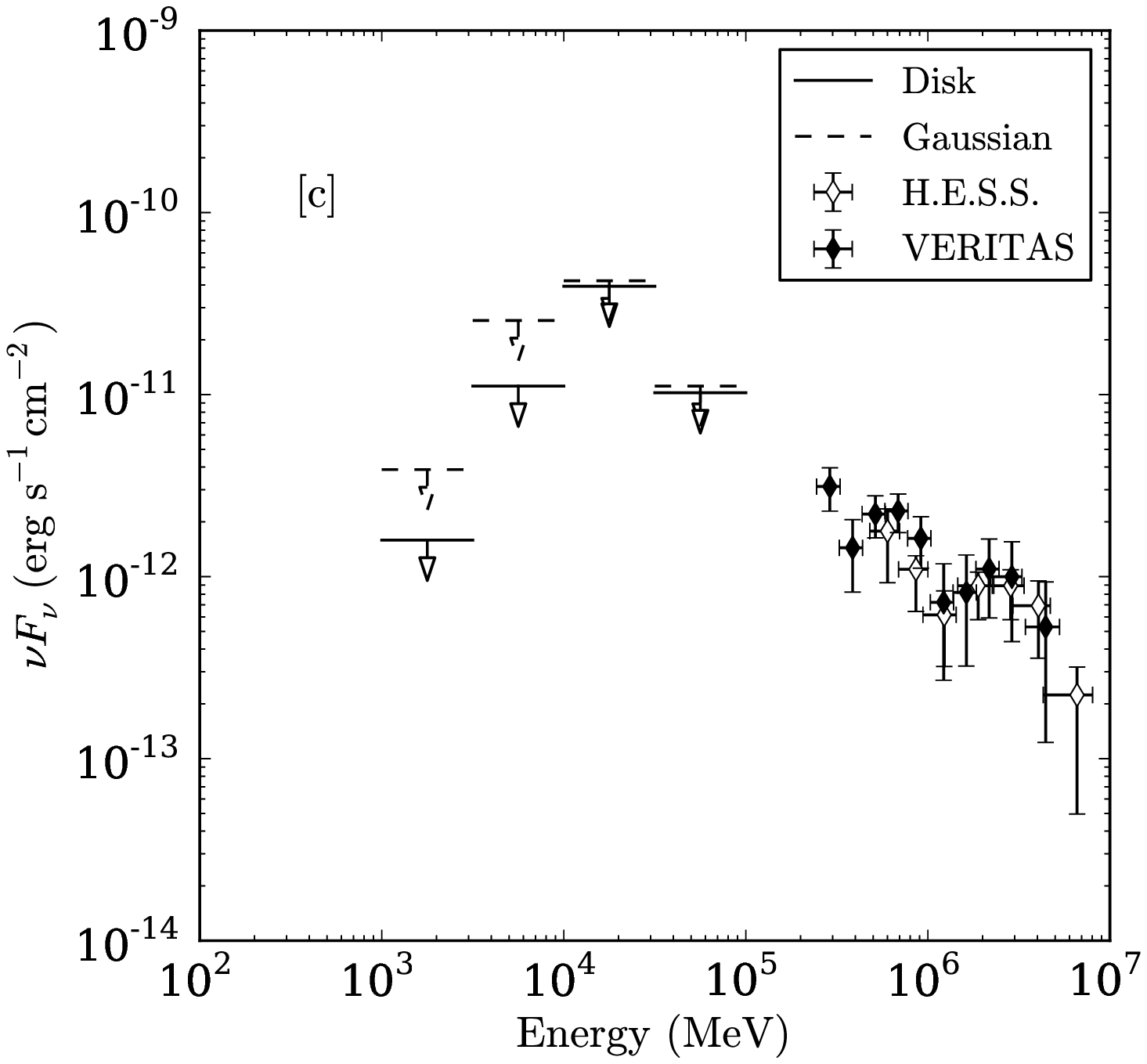} 
\caption{Upper limits on the energy flux from 1ES0229+200 at the 95\% confidence level assuming a 0.1$^\circ$ (\textbf{a}), 0.5$^\circ$ (\textbf{b}), and 1.0$^\circ$ (\textbf{c}) halo, plotted with observations from HESS and VERITAS. 
}
\label{FIGURE::1es0229}
\end{figure}
\begin{figure}[htb]
\centering
    \includegraphics[width=\twothirdscolfigwidth]{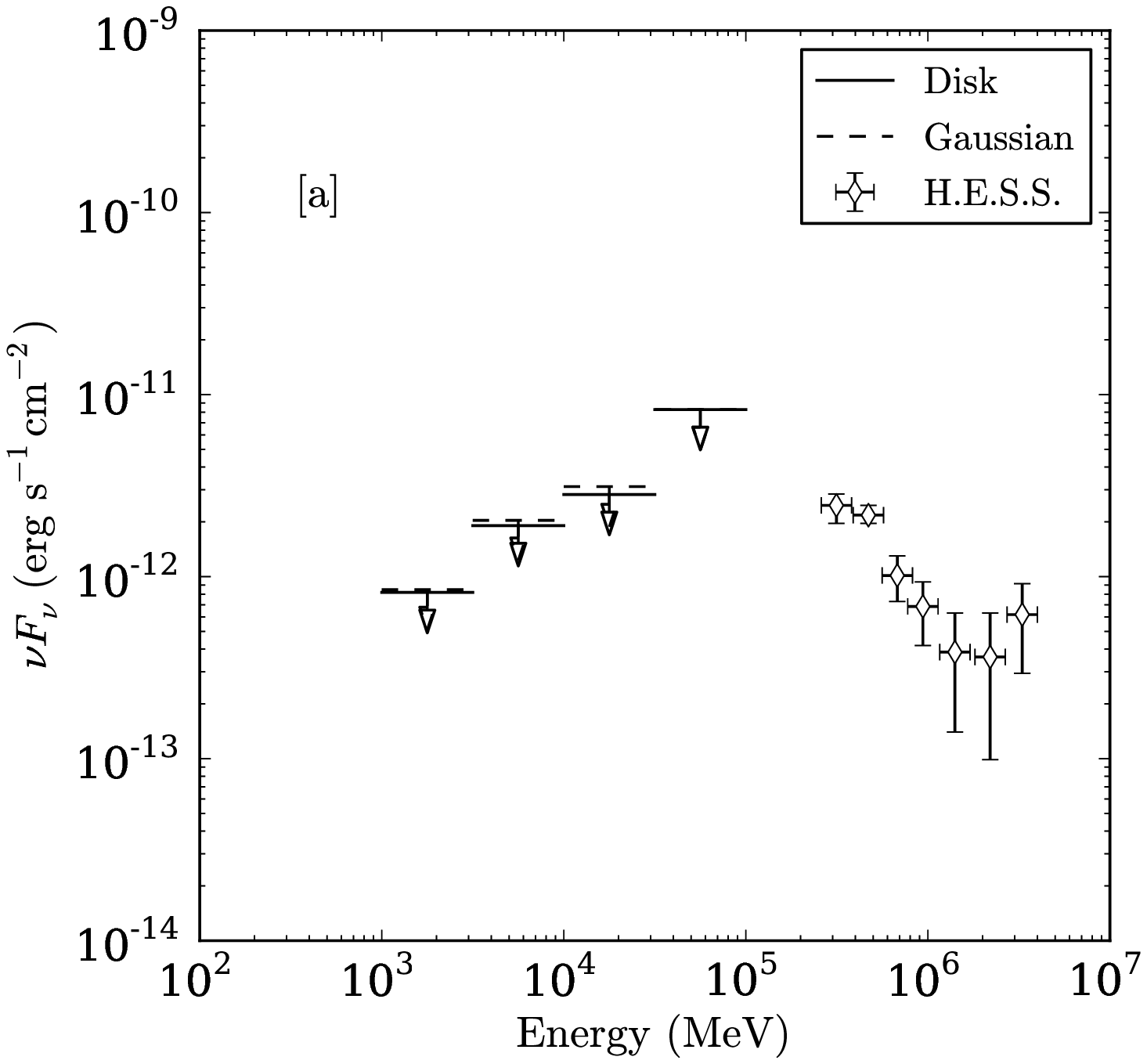}\hfill%
    \includegraphics[width=\twothirdscolfigwidth]{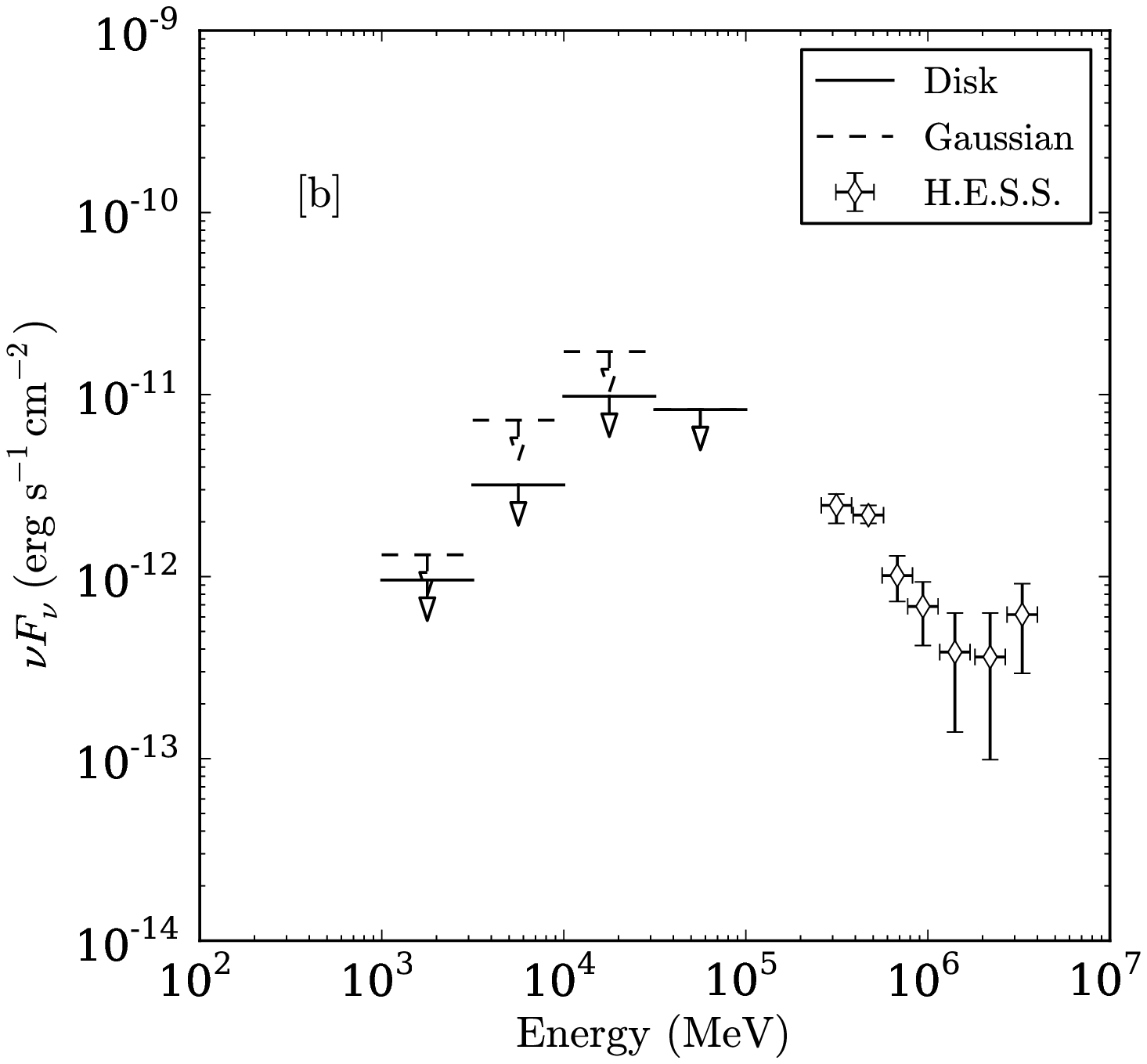}\hfill%
    \includegraphics[width=\twothirdscolfigwidth]{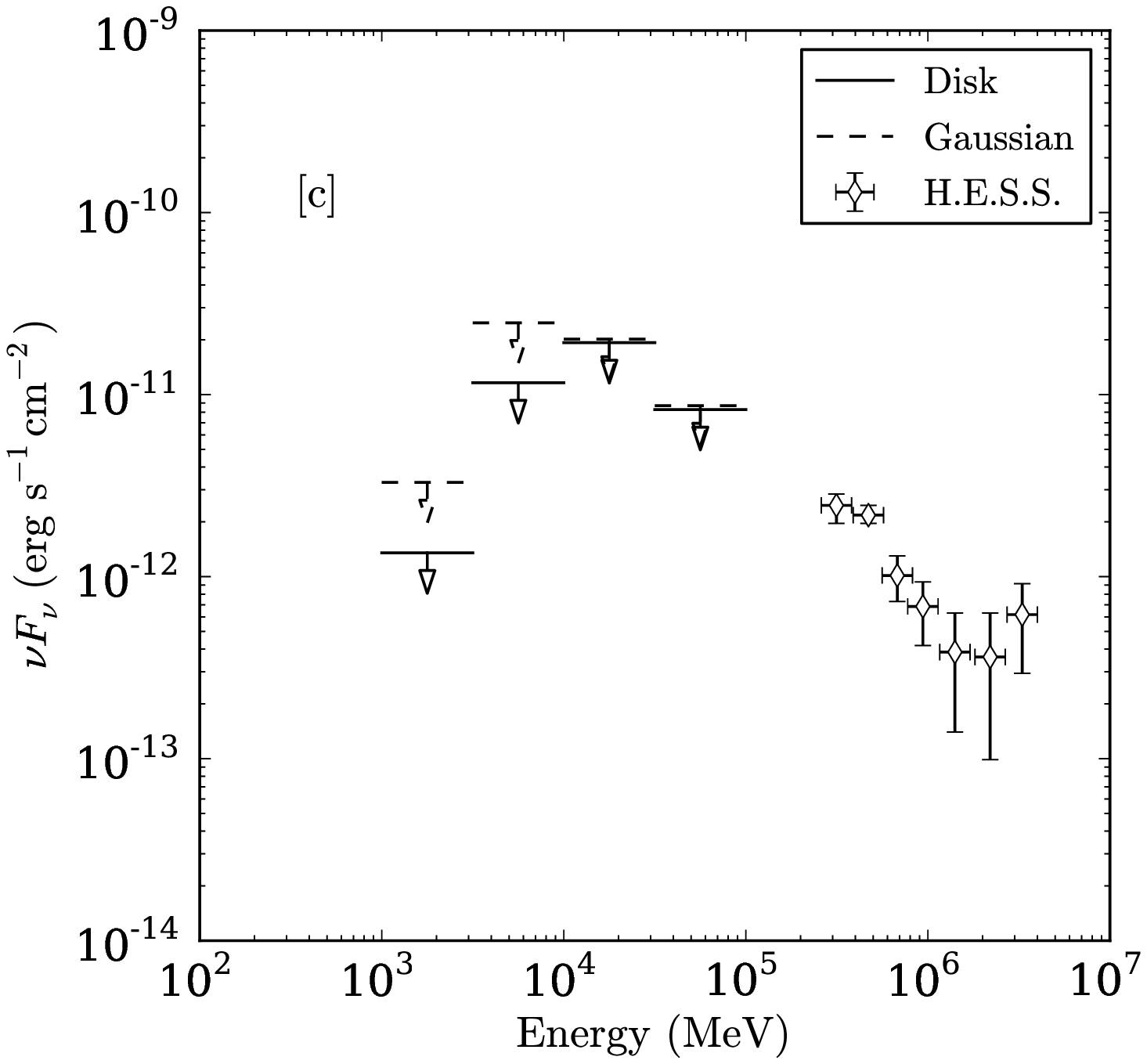} 
\caption{Upper limits on the energy flux from 1ES0347-121 at the 95\% confidence level, plotted with observations from HESS assuming a 0.1$^\circ$ (\textbf{a}), 0.5$^\circ$ (\textbf{b}), and 1.0$^\circ$ (\textbf{c}) halo.}
\label{FIGURE::1es0347}
\end{figure}

There were no detections for $TS>4~(S>2)$ in any energy band for
either source, so we calculated the 95\% upper limits on $N^{halo}$
for each source in each energy band.  We converted the 95\% upper
limits into flux measurements by calculating the two-year exposure for
each source, giving the results plotted in Figures
\ref{FIGURE::1es0229} and
\ref{FIGURE::1es0347} 
with H.E.S.S. and VERITAS measurements \citep{REF:2007A&A...475L...9A,
  REF:2010HEAD...11.3318P, REF:2007A&A...473L..25A}.  We find good
agreement between our upper limits and those derived by
\citet{REF:2010.Neronov2}, \citet{REF:2010.Tavecchio}, and
\citet{REF:2011.DermerHalo}.


\section{Discussion and Summary}



As noted in the Section 1, TeV photons emitted by blazars can generate
a secondary cascade when they pair produce on photons of the EBL and
create electron-positron pairs.  These pairs subsequently upscatter
CMB photons through the inverse Compton process generating a secondary
component of gamma-ray emission at GeV energies.  By deflecting the
trajectories of off-axis pairs into our line of sight, the IGMF can
create a halo of cascade radiation around the blazar.  The angular
profile of this halo changes with the strength of the IGMF and would
appear point-like within the angular resolution limit of the Fermi-LAT
if the magnetic field strength is small ($\lesssim 10^{-18} $ G).
In the limit of large magnetic field strength, the cascade radiation
has a maximum angular size determined by the surface subtended by the
jet at a distance $\tau \simeq$ 1 from the source where $\tau$ is the
pair-production optical depth of the primary VHE gamma-rays
\citep{2010ApJ...719L.130N,REF:2010.Tavecchio}.
However, several other processes are known that could form extended
emission around blazars such as synchrotron radiation from leptonic
secondaries of ultra-high energy cosmic ray (UHECR) protons undergoing
photopion processes in $\gg 10^{-12}$ G fields
\citep{2007Ap&SS.309..465G} or cascade radiation induced by photopair
losses of UHECRs \citep{2010APh....33...81E,2010PhRvL.104n1102E}.  Nor
do halos provide the only opportunity to measure $B_{IGMF}$. Combined
GeV-TeV spectral analysis of moderate redshift sources ($z\approx
0.2$) has already been used to infer values of $B_{IGMF}\gtrsim
10^{-15}$~G under the assumption of persistent blazar emission over
long times \citep{REF:2010.Neronov2,REF:2010.Tavecchio}.  For blazars
radiating for only a few years at a constant flux level,
$B_{IGMF}\gtrsim 10^{-17}$~G
\citep{2011ApJ...727L...4D,2011A&A...529A.144T} or $B_{IGMF}\gtrsim
10^{-18}$~G \citep{REF:2011.DermerHalo} depending on the assumed
unabsorbed spectrum.  Searches for delayed secondary radiation from
impulsive $\gamma$-ray sources provide another technique to constrain
$B_{IGMF}$ \citep[e.g.,][]{1995Natur.374..430P,2008ApJ...686L..67M}.

Claims for the existence of halos around 170 stacked hard-spectrum AGN
in the 1LAC were made by \citet{REF:2010.AndoKus} using the
\irf{P6\_V3} \textit{Diffuse} PSF and a halo component,
$dN/d\Omega(\theta) \propto \exp{\left(-\theta^4\right)}$.  This
angular model is broader than the Disk model, but narrower than the
Gaussian model, making it a reasonable comparison to our procedure.
We attribute the detection of apparent extended emission of AGN by
Ando and Kusenko to the difference between the \irf{P6\_V3} PSFs and
the actual PSF as inferred here from flight data.  Our results are
consistent with \citet{REF:2011.Neronov1} who found no evidence in the
{\it Fermi}-LAT data for extended halo emission.
\citet{REF:2010.AndoKus} further claimed a detection of a halo
component in the 3--10 GeV energy range using a low-redshift AGN
subsample and the Crab nebula as a PSF calibration source.  In the
same energy range our analysis, which used Vela and Geminga as
calibration sources, set an 95\% C. L. upper limits of 0.05 and 0.11 on
the fraction of halo emission in a sample of low-redshift AGN for a
Disk and Gaussian halo models, respectively (see Table
\ref{TABLE::lzsum}). 
Furthermore no significant detections of a halo component were found
for the range of halo sizes and shapes tested.

In conclusion, we have derived the PSF through on-orbit data (\irf{P6\_V11}), revealing
that the MC PSF (\irf{P6\_V3}) significantly underestimates the 68\%
containment radius of the PSF at GeV energies.
The discrepancies are larger for back- than
front-converting events with the underestimate of the 68\%
containment radius at 5--10 GeV reaching 25\% and 50\%
for the two event classes, respectively.  The \irf{P6\_V11} PSF
provides a better representation of the 68\% containment radius at
high energies.  However, due to the limitations of the single-King-function 
parameterization that we used for \irf{P6\_V11}, the \irf{P6\_V3} PSF more
accurately describes the 95\% containment radius up to $\sim$7 GeV.  Furthermore, the
\irf{P6\_V11} PSF was derived from an event sample which ignored the inclination 
angle of the incident $\gamma$ ray, and therefore does not model the dependence of the PSF on the inclination angle. 
The improved PSF, \irf{P6\_V11}, 
supersedes the \irf{P6\_V3} PSF 
that has a systematically narrower model than the distributions of 
$\gamma$ rays around bright sources. 
Upper limits were derived for the flux of an extended halo component in
analyses of stacked AGN.  No evidence for halos around extragalactic
TeV sources is found in our analysis, consistent with the limits found
from other recent studies.

\acknowledgements
The \textit{Fermi} LAT Collaboration acknowledges generous ongoing support
from a number of agencies and institutes that have supported both the
development and the operation of the LAT as well as scientific data analysis.
These include the National Aeronautics and Space Administration and the
Department of Energy in the United States, the Commissariat \`a l'Energie
Atomique and the Centre National de la Recherche Scientifique / Institut
National de Physique Nucl\'eaire et de Physique des Particules in France,
the Agenzia Spaziale Italiana and the Istituto Nazionale di Fisica Nucleare in
Italy, the Ministry of Education, Culture, Sports, Science and Technology
(MEXT), High Energy Accelerator Research Organization (KEK) and Japan Aerospace
Exploration Agency (JAXA) in Japan, and the K.~A.~Wallenberg Foundation,
the Swedish Research Council and the Swedish National Space Board in Sweden.

Additional support for science analysis during the operations phase is
gratefully acknowledged from the Istituto Nazionale di Astrofisica in Italy and
the Centre National d'\'Etudes Spatiales in France.

The Parkes radio telescope is part of the Australia Telescope which is funded by the Commonwealth Government for operation as a National Facility managed by CSIRO. 
We thank our colleagues for their assistance with the radio timing observations.

\end{doublespace}
\bibliography{halopaper}

\begin{thebibliography}{41}
\expandafter\ifx\csname natexlab\endcsname\relax\def\natexlab#1{#1}\fi

\bibitem[{{Abdo} {et~al.}(2009){Abdo}, {Ackermann}, {Ajello}, {Ampe},
  {Anderson}, {Atwood}, {Axelsson}, {Bagagli}, {Baldini}, {Ballet}, \&
  et~al.}]{REF:2009.OnOrbitCalib}
{Abdo}, A.~A., {et~al.} 2009, Astroparticle Physics, 32, 193

\bibitem[{{Abdo} {et~al.}(2010{\natexlab{a}}){Abdo}, {Ackermann}, {Ajello},
  {Allafort}, {Antolini}, {Atwood}, {Axelsson}, {Baldini}, {Ballet},
  {Barbiellini}, \& et~al.}]{REF:2010.1FGL}
---. 2010{\natexlab{a}}, \apjs, 188, 405

\bibitem[{{Abdo} {et~al.}(2010{\natexlab{b}}){Abdo}, {Ackermann}, {Ajello},
  {Allafort}, {Baldini}, {Ballet}, {Barbiellini}, {Bastieri}, {Bechtol},
  {Bellazzini}, {Berenji}, {Blandford}, {Bloom}, {Bonamente}, {Borgland},
  {Bouvier}, {Bregeon}, {Brez}, {Brigida}, {Bruel}, {Burnett}, {Buson},
  {Caliandro}, {Cameron}, {Caraveo}, {Carrigan}, {Casandjian}, {Cecchi}, {{\c
  C}elik}, {Chekhtman}, {Chung}, {Chiang}, {Ciprini}, {Claus}, {Cohen-Tanugi},
  {Conrad}, {de Angelis}, {de Palma}, {Dormody}, {Silva}, {Drell}, {Dubois},
  {Dumora}, {Farnier}, {Favuzzi}, {Fegan}, {Focke}, {Fortin}, {Frailis},
  {Fukazawa}, {Funk}, {Fusco}, {Gargano}, {Gehrels}, {Germani}, {Giavitto},
  {Giglietto}, {Giordano}, {Glanzman}, {Godfrey}, {Grenier}, {Grondin},
  {Grove}, {Guillemot}, {Guiriec}, {Harding}, {Hayashida}, {Hays}, {Horan},
  {Hughes}, {Jackson}, {J{\'o}hannesson}, {Johnson}, {Johnson}, {Johnson},
  {Johnston}, {Kamae}, {Katagiri}, {Kataoka}, {Kawai}, {Kerr},
  {Kn{\"o}dlseder}, {Kuss}, {Lande}, {Latronico}, {Lee}, {Lemoine-Goumard},
  {Llena Garde}, {Longo}, {Loparco}, {Lott}, {Lovellette}, {Lubrano}, {Makeev},
  {Marelli}, {Mazziotta}, {McEnery}, {Meurer}, {Michelson}, {Mitthumsiri},
  {Mizuno}, {Moiseev}, {Monte}, {Monzani}, {Morselli}, {Moskalenko}, {Murgia},
  {Nakamori}, {Nolan}, {Norris}, {Noutsos}, {Nuss}, {Ohsugi}, {Omodei},
  {Orlando}, {Ormes}, {Ozaki}, {Paneque}, {Panetta}, {Parent}, {Pelassa},
  {Pepe}, {Pesce-Rollins}, {Pierbattista}, {Piron}, {Porter}, {Rain{\`o}},
  {Rando}, {Ray}, {Rea}, {Reimer}, {Reimer}, {Reposeur}, {Ritz}, {Rodriguez},
  {Romani}, {Roth}, {Ryde}, {Sadrozinski}, {Sanchez}, {Sander}, {Saz
  Parkinson}, {Scargle}, {Sgr{\`o}}, {Siskind}, {Smith}, {Smith}, {Spandre},
  {Spinelli}, {Strickman}, {Suson}, {Tajima}, {Takahashi}, {Takahashi},
  {Tanaka}, {Thayer}, {Thayer}, {Thompson}, {Tibaldo}, {Torres}, {Tosti},
  {Tramacere}, {Uchiyama}, {Usher}, {Van Etten}, {Vasileiou}, {Venter},
  {Vilchez}, {Vitale}, {Waite}, {Wang}, {Weltevrede}, {Winer}, {Wood},
  {Ylinen}, \& {Ziegler}}]{REF:2010.VelaPWN}
---. 2010{\natexlab{b}}, \apj, 713, 146

\bibitem[{{Abdo} {et~al.}(2010{\natexlab{c}}){Abdo}, {Ackermann}, {Ajello},
  {Baldini}, {Ballet}, {Barbiellini}, {Bastieri}, {Baughman}, {Bechtol},
  {Bellazzini}, {Berenji}, {Bignami}, {Blandford}, {Bloom}, {Bonamente},
  {Borgland}, {Bregeon}, {Brez}, {Brigida}, {Bruel}, {Burnett}, {Caliandro},
  {Cameron}, {Caraveo}, {Casandjian}, {Cecchi}, {{\c C}elik}, {Charles},
  {Chekhtman}, {Cheung}, {Chiang}, {Ciprini}, {Claus}, {Cohen-Tanugi},
  {Conrad}, {Dermer}, {de Palma}, {Dormody}, {Silva}, {Drell}, {Dubois},
  {Dumora}, {Edmonds}, {Farnier}, {Favuzzi}, {Fegan}, {Focke}, {Fortin},
  {Frailis}, {Fukazawa}, {Funk}, {Fusco}, {Gargano}, {Gasparrini}, {Gehrels},
  {Germani}, {Giavitto}, {Giglietto}, {Giordano}, {Glanzman}, {Godfrey},
  {Grenier}, {Grondin}, {Grove}, {Guillemot}, {Guiriec}, {Hadasch}, {Harding},
  {Hays}, {Hughes}, {J{\'o}hannesson}, {Johnson}, {Johnson}, {Johnson},
  {Kamae}, {Katagiri}, {Kataoka}, {Kawai}, {Kerr}, {Kn{\"o}dlseder}, {Kuss},
  {Lande}, {Latronico}, {Lemoine-Goumard}, {Longo}, {Loparco}, {Lott},
  {Lovellette}, {Lubrano}, {Makeev}, {Marelli}, {Mazziotta}, {McEnery},
  {Meurer}, {Michelson}, {Mitthumsiri}, {Mizuno}, {Moiseev}, {Monte},
  {Monzani}, {Morselli}, {Moskalenko}, {Murgia}, {Nolan}, {Norris}, {Nuss},
  {Ohsugi}, {Omodei}, {Orlando}, {Ormes}, {Ozaki}, {Paneque}, {Panetta},
  {Parent}, {Pelassa}, {Pepe}, {Pesce-Rollins}, {Piron}, {Porter}, {Rain{\`o}},
  {Rando}, {Ray}, {Razzano}, {Reimer}, {Reimer}, {Reposeur}, {Rochester},
  {Rodriguez}, {Romani}, {Roth}, {Ryde}, {Sadrozinski}, {Sander}, {Saz
  Parkinson}, {Scargle}, {Sgr{\`o}}, {Siskind}, {Smith}, {Smith}, {Spandre},
  {Spinelli}, {Strickman}, {Suson}, {Takahashi}, {Takahashi}, {Tanaka},
  {Thayer}, {Thayer}, {Thompson}, {Tibaldo}, {Torres}, {Tosti}, {Tramacere},
  {Usher}, {Van Etten}, {Vasileiou}, {Venter}, {Vilchez}, {Vitale}, {Waite},
  {Wang}, {Watters}, {Winer}, {Wood}, {Ylinen}, \&
  {Ziegler}}]{REF:2010.Geminga}
---. 2010{\natexlab{c}}, \apj, 720, 272

\bibitem[{{Abdo} {et~al.}(2010{\natexlab{d}}){Abdo}, {Ackermann}, {Ajello},
  {Allafort}, {Antolini}, {Atwood}, {Axelsson}, {Baldini}, {Ballet},
  {Barbiellini}, {Bastieri}, {Baughman}, {Bechtol}, {Bellazzini}, {Berenji},
  {Blandford}, {Bloom}, {Bogart}, {Bonamente}, {Borgland}, {Bouvier},
  {Bregeon}, {Brez}, {Brigida}, {Bruel}, {Buehler}, {Burnett}, {Buson},
  {Caliandro}, {Cameron}, {Cannon}, {Caraveo}, {Carrigan}, {Casandjian},
  {Cavazzuti}, {Cecchi}, {{\c C}elik}, {Celotti}, {Charles}, {Chekhtman},
  {Chen}, {Cheung}, {Chiang}, {Ciprini}, {Claus}, {Cohen-Tanugi}, {Conrad},
  {Costamante}, {Cotter}, {Cutini}, {D'Elia}, {Dermer}, {de Angelis}, {de
  Palma}, {De Rosa}, {Digel}, {Silva}, {Drell}, {Dubois}, {Dumora}, {Escande},
  {Farnier}, {Favuzzi}, {Fegan}, {Ferrara}, {Focke}, {Fortin}, {Frailis},
  {Fukazawa}, {Funk}, {Fusco}, {Gargano}, {Gasparrini}, {Gehrels}, {Germani},
  {Giebels}, {Giglietto}, {Giommi}, {Giordano}, {Giroletti}, {Glanzman},
  {Godfrey}, {Grandi}, {Grenier}, {Grondin}, {Grove}, {Guiriec}, {Hadasch},
  {Harding}, {Hayashida}, {Hays}, {Healey}, {Hill}, {Horan}, {Hughes},
  {Iafrate}, {Itoh}, {J{\'o}hannesson}, {Johnson}, {Johnson}, {Johnson},
  {Johnson}, {Kamae}, {Katagiri}, {Kataoka}, {Kawai}, {Kerr}, {Kn{\"o}dlseder},
  {Kuss}, {Lande}, {Latronico}, {Lavalley}, {Lemoine-Goumard}, {Llena Garde},
  {Longo}, {Loparco}, {Lott}, {Lovellette}, {Lubrano}, {Madejski}, {Makeev},
  {Malaguti}, {Massaro}, {Mazziotta}, {McConville}, {McEnery}, {McGlynn},
  {Michelson}, {Mitthumsiri}, {Mizuno}, {Moiseev}, {Monte}, {Monzani},
  {Morselli}, {Moskalenko}, {Murgia}, {Nolan}, {Norris}, {Nuss}, {Ohno},
  {Ohsugi}, {Omodei}, {Orlando}, {Ormes}, {Ozaki}, {Paneque}, {Panetta},
  {Parent}, {Pelassa}, {Pepe}, {Pesce-Rollins}, {Piranomonte}, {Piron},
  {Porter}, {Rain{\`o}}, {Rando}, {Razzano}, {Reimer}, {Reimer}, {Reposeur},
  {Ripken}, {Ritz}, {Rodriguez}, {Romani}, {Roth}, {Ryde}, {Sadrozinski},
  {Sanchez}, {Sander}, {Saz Parkinson}, {Scargle}, {Sgr{\`o}}, {Shaw},
  {Siskind}, {Smith}, {Spandre}, {Spinelli}, {Starck}, {Stawarz}, {Strickman},
  {Suson}, {Tajima}, {Takahashi}, {Takahashi}, {Tanaka}, {Taylor}, {Thayer},
  {Thayer}, {Thompson}, {Tibaldo}, {Torres}, {Tosti}, {Tramacere}, {Ubertini},
  {Uchiyama}, {Usher}, {Vasileiou}, {Vilchez}, {Villata}, {Vitale}, {Waite},
  {Wallace}, {Wang}, {Winer}, {Wood}, {Yang}, {Ylinen}, \&
  {Ziegler}}]{REF:2010ApJ...715..429A}
---. 2010{\natexlab{d}}, \apj, 715, 429

\bibitem[{{Ackermann} {et~al.}(2011){Ackermann}, {Ajello}, {Baldini}, {Ballet},
  {Barbiellini}, {Bastieri}, {Bechtol}, {Bellazzini}, {Berenji}, {Bloom},
  {Bonamente}, {Borgland}, {Bouvier}, {Bregeon}, {Brez}, {Brigida}, {Bruel},
  {Buehler}, {Buson}, {Caliandro}, {Cameron}, {Camilo}, {Caraveo},
  {Casandjian}, {Cecchi}, {{\c C}elik}, {Charles}, {Chekhtman}, {Cheung},
  {Chiang}, {Ciprini}, {Claus}, {Cognard}, {Cohen-Tanugi}, {Conrad}, {Dermer},
  {de Angelis}, {de Luca}, {de Palma}, {Digel}, {Silva}, {Drell}, {Dubois},
  {Dumora}, {Favuzzi}, {Focke}, {Frailis}, {Fukazawa}, {Funk}, {Fusco},
  {Gargano}, {Germani}, {Giglietto}, {Giommi}, {Giordano}, {Giroletti},
  {Glanzman}, {Godfrey}, {Grenier}, {Grondin}, {Grove}, {Guillemot}, {Guiriec},
  {Hadasch}, {Hanabata}, {Harding}, {Hayashi}, {Hays}, {Hobbs}, {Hughes},
  {J{\'o}hannesson}, {Johnson}, {Johnson}, {Johnston}, {Kamae}, {Katagiri},
  {Kataoka}, {Keith}, {Kerr}, {Kn{\"o}dlseder}, {Kramer}, {Kuss}, {Lande},
  {Latronico}, {Lee}, {Lemoine-Goumard}, {Longo}, {Loparco}, {Lovellette},
  {Lubrano}, {Lyne}, {Makeev}, {Marelli}, {Mazziotta}, {McEnery}, {Mehault},
  {Michelson}, {Mizuno}, {Moiseev}, {Monte}, {Monzani}, {Morselli},
  {Moskalenko}, {Murgia}, {Nakamori}, {Naumann-Godo}, {Nolan}, {Noutsos},
  {Nuss}, {Ohsugi}, {Okumura}, {Ormes}, {Paneque}, {Panetta}, {Parent},
  {Pelassa}, {Pepe}, {Pesce-Rollins}, {Piron}, {Porter}, {Rain{\`o}}, {Rando},
  {Ransom}, {Ray}, {Razzano}, {Rea}, {Reimer}, {Reimer}, {Reposeur}, {Ripken},
  {Ritz}, {Romani}, {Sadrozinski}, {Sander}, {Saz Parkinson}, {Sgr{\`o}},
  {Siskind}, {Smith}, {Smith}, {Spandre}, {Spinelli}, {Strickman}, {Suson},
  {Takahashi}, {Takahashi}, {Tanaka}, {Thayer}, {Thayer}, {Theureau},
  {Thompson}, {Thorsett}, {Tibaldo}, {Torres}, {Tosti}, {Tramacere},
  {Uchiyama}, {Uehara}, {Usher}, {Vandenbroucke}, {Van Etten}, {Vasileiou},
  {Vilchez}, {Vitale}, {Waite}, {Wang}, {Weltevrede}, {Winer}, {Wood}, {Yang},
  {Ylinen}, \& {Ziegler}}]{2011ApJ...726...35A}
{Ackermann}, M., {et~al.} 2011, \apj, 726, 35

\bibitem[{{Ackermann} {et~al.}(2012){Ackermann}, {Ajello}, {Albert},
  {Allafort}, {Atwood}, {Axelsson}, {Baldini}, {Ballet}, {Barbiellini},
  {Bastieri}, {Bechtol}, {Bellazzini}, {Bissaldi}, {Blandford}, {Bloom},
  {Bogart}, {Bonamente}, {Borgland}, {Bottacini}, {Bouvier}, {Brandt},
  {Bregeon}, {Brigida}, {Bruel}, {Buehler}, {Burnett}, {Buson}, {Caliandro},
  {Cameron}, {Caraveo}, {Casandjian}, {Cavazzuti}, {Cecchi}, {{\c C}elik},
  {Charles}, {Chaves}, {Chekhtman}, {Cheung}, {Chiang}, {Ciprini}, {Claus},
  {Cohen-Tanugi}, {Conrad}, {Corbet}, {Cutini}, {D'Ammando}, {Davis}, {de
  Angelis}, {DeKlotz}, {de Palma}, {Dermer}, {Digel}, {Silva}, {Drell},
  {Drlica-Wagner}, {Dubois}, {Favuzzi}, {Fegan}, {Ferrara}, {Focke}, {Fortin},
  {Fukazawa}, {Funk}, {Fusco}, {Gargano}, {Gasparrini}, {Gehrels}, {Giebels},
  {Giglietto}, {Giordano}, {Giroletti}, {Glanzman}, {Godfrey}, {Grenier},
  {Grove}, {Guiriec}, {Hadasch}, {Hayashida}, {Hays}, {Horan}, {Hou}, {Hughes},
  {Jackson}, {Jogler}, {J{\'o}hannesson}, {Johnson}, {Johnson}, {Johnson},
  {Kamae}, {Katagiri}, {Kataoka}, {Kerr}, {Kn{\"o}dlseder}, {Kuss}, {Lande},
  {Larsson}, {Latronico}, {Lavalley}, {Lemoine-Goumard}, {Longo}, {Loparco},
  {Lott}, {Lovellette}, {Lubrano}, {Mazziotta}, {McConville}, {McEnery},
  {Mehault}, {Michelson}, {Mitthumsiri}, {Mizuno}, {Moiseev}, {Monte},
  {Monzani}, {Morselli}, {Moskalenko}, {Murgia}, {Naumann-Godo}, {Nemmen},
  {Nishino}, {Norris}, {Nuss}, {Ohno}, {Ohsugi}, {Okumura}, {Omodei},
  {Orienti}, {Orlando}, {Ormes}, {Paneque}, {Panetta}, {Perkins},
  {Pesce-Rollins}, {Pierbattista}, {Piron}, {Pivato}, {Porter}, {Racusin},
  {Rain{\`o}}, {Rando}, {Razzano}, {Razzaque}, {Reimer}, {Reimer}, {Reposeur},
  {Reyes}, {Ritz}, {Rochester}, {Romoli}, {Roth}, {Sadrozinski}, {Sanchez},
  {Saz Parkinson}, {Sbarra}, {Scargle}, {Sgr{\`o}}, {Siegal-Gaskins},
  {Siskind}, {Spandre}, {Spinelli}, {Stephens}, {Suson}, {Tajima}, {Takahashi},
  {Tanaka}, {Thayer}, {Thayer}, {Thompson}, {Tibaldo}, {Tinivella}, {Tosti},
  {Troja}, {Usher}, {Vandenbroucke}, {Van Klaveren}, {Vasileiou}, {Vianello},
  {Vitale}, {Waite}, {Wallace}, {Winer}, {Wood}, {Wood}, {Wood}, {Yang}, \&
  {Zimmer}}]{2012ApJS..203....4A}
---. 2012, \apjs, 203, 4

\bibitem[{{Aharonian} {et~al.}(2007{\natexlab{a}}){Aharonian}, {Akhperjanian},
  {Barres de Almeida}, {Bazer-Bachi}, {Behera}, {Beilicke}, {Benbow},
  {Bernl{\"o}hr}, {Boisson}, {Bolz}, {Borrel}, {Braun}, {Brion}, {Brown},
  {B{\"u}hler}, {Bulik}, {B{\"u}sching}, {Boutelier}, {Carrigan}, {Chadwick},
  {Chounet}, {Clapson}, {Coignet}, {Cornils}, {Costamante}, {Dalton},
  {Degrange}, {Dickinson}, {Djannati-Ata{\"i}}, {Domainko}, {O'C.~Drury},
  {Dubois}, {Dubus}, {Dyks}, {Egberts}, {Emmanoulopoulos}, {Espigat},
  {Farnier}, {Feinstein}, {Fiasson}, {F{\"o}rster}, {Fontaine}, {Funk},
  {F{\"u}{\ss}ling}, {Gallant}, {Giebels}, {Glicenstein}, {Gl{\"u}ck}, {Goret},
  {Hadjichristidis}, {Hauser}, {Hauser}, {Heinzelmann}, {Henri}, {Hermann},
  {Hinton}, {Hoffmann}, {Hofmann}, {Holleran}, {Hoppe}, {Horns},
  {Jacholkowska}, {de Jager}, {Jung}, {Katarzy{\'n}ski}, {Kendziorra},
  {Kerschhaggl}, {Kh{\'e}lifi}, {Keogh}, {Komin}, {Kosack}, {Lamanna},
  {Latham}, {Lemi{\`e}re}, {Lemoine-Goumard}, {Lenain}, {Lohse}, {Martin},
  {Martineau-Huynh}, {Marcowith}, {Masterson}, {Maurin}, {Maurin}, {McComb},
  {Moderski}, {Moulin}, {de Naurois}, {Nedbal}, {Nolan}, {Ohm}, {Olive}, {de
  O{\~n}a Wilhelmi}, {Orford}, {Osborne}, {Ostrowski}, {Panter}, {Pedaletti},
  {Pelletier}, {Petrucci}, {Pita}, {P{\"u}hlhofer}, {Punch}, {Ranchon},
  {Raubenheimer}, {Raue}, {Rayner}, {Renaud}, {Ripken}, {Rob}, {Rolland},
  {Rosier-Lees}, {Rowell}, {Rudak}, {Ruppel}, {Sahakian}, {Santangelo},
  {Schlickeiser}, {Sch{\"o}ck}, {Schr{\"o}der}, {Schwanke}, {Schwarzburg},
  {Schwemmer}, {Shalchi}, {Sol}, {Spangler}, {Stawarz}, {Steenkamp},
  {Stegmann}, {Superina}, {Tam}, {Tavernet}, {Terrier}, {van Eldik},
  {Vasileiadis}, {Venter}, {Vialle}, {Vincent}, {Vivier}, {V{\"o}lk}, {Volpe},
  {Wagner}, {Ward}, {Zdziarski}, \& {Zech}}]{REF:2007A&A...473L..25A}
{Aharonian}, F., {et~al.} 2007{\natexlab{a}}, \aap, 473, L25

\bibitem[{{Aharonian} {et~al.}(2007{\natexlab{b}}){Aharonian}, {Akhperjanian},
  {Barres de Almeida}, {Bazer-Bachi}, {Behera}, {Beilicke}, {Benbow},
  {Bernl{\"o}hr}, {Boisson}, {Bolz}, {Borrel}, {Braun}, {Brion}, {Brown},
  {B{\"u}hler}, {Bulik}, {B{\"u}sching}, {Boutelier}, {Carrigan}, {Chadwick},
  {Chounet}, {Clapson}, {Coignet}, {Cornils}, {Costamante}, {Dalton},
  {Degrange}, {Dickinson}, {Djannati-Ata{\"i}}, {Domainko}, {O'C.~Drury},
  {Dubois}, {Dubus}, {Dyks}, {Egberts}, {Emmanoulopoulos}, {Espigat},
  {Farnier}, {Feinstein}, {Fiasson}, {F{\"o}rster}, {Fontaine}, {Funk},
  {F{\"u}{\ss}ling}, {Gallant}, {Giebels}, {Glicenstein}, {Gl{\"u}ck}, {Goret},
  {Hadjichristidis}, {Hauser}, {Hauser}, {Heinzelmann}, {Henri}, {Hermann},
  {Hinton}, {Hoffmann}, {Hofmann}, {Holleran}, {Hoppe}, {Horns},
  {Jacholkowska}, {de Jager}, {Jung}, {Katarzy{\'n}ski}, {Kendziorra},
  {Kerschhaggl}, {Kh{\'e}lifi}, {Keogh}, {Komin}, {Kosack}, {Lamanna},
  {Latham}, {Lemi{\`e}re}, {Lemoine-Goumard}, {Lenain}, {Lohse}, {Martin},
  {Martineau-Huynh}, {Marcowith}, {Masterson}, {Maurin}, {Maurin}, {McComb},
  {Moderski}, {Moulin}, {de Naurois}, {Nedbal}, {Nolan}, {Ohm}, {Olive}, {de
  O{\~n}a Wilhelmi}, {Orford}, {Osborne}, {Ostrowski}, {Panter}, {Pedaletti},
  {Pelletier}, {Petrucci}, {Pita}, {P{\"u}hlhofer}, {Punch}, {Ranchon},
  {Raubenheimer}, {Raue}, {Rayner}, {Renaud}, {Ripken}, {Rob}, {Rolland},
  {Rosier-Lees}, {Rowell}, {Rudak}, {Ruppel}, {Sahakian}, {Santangelo},
  {Schlickeiser}, {Sch{\"o}ck}, {Schr{\"o}der}, {Schwanke}, {Schwarzburg},
  {Schwemmer}, {Shalchi}, {Sol}, {Spangler}, {Stawarz}, {Steenkamp},
  {Stegmann}, {Superina}, {Tam}, {Tavernet}, {Terrier}, {van Eldik},
  {Vasileiadis}, {Venter}, {Vialle}, {Vincent}, {Vivier}, {V{\"o}lk}, {Volpe},
  {Wagner}, {Ward}, {Zdziarski}, \& {Zech}}]{REF:2007A&A...475L...9A}
---. 2007{\natexlab{b}}, \aap, 475, L9

\bibitem[{{Aleksi{\'c}} {et~al.}(2010){Aleksi{\'c}}, {Antonelli}, {Antoranz},
  {Backes}, {Baixeras}, {Barrio}, {Bastieri}, {Becerra Gonz{\'a}lez},
  {Bednarek}, {Berdyugin}, {Berger}, {Bernardini}, {Biland}, {Blanch}, {Bock},
  {Bonnoli}, {Bordas}, {Borla Tridon}, {Bosch-Ramon}, {Bose}, {Braun}, {Bretz},
  {Britzger}, {Camara}, {Carmona}, {Carosi}, {Colin}, {Commichau}, {Contreras},
  {Cortina}, {Costado}, {Covino}, {Dazzi}, {de Angelis}, {de Cea Del Pozo}, {de
  Los Reyes}, {de Lotto}, {de Maria}, {de Sabata}, {Delgado Mendez}, {Doert},
  {Dom{\'{\i}}nguez}, {Dominis Prester}, {Dorner}, {Doro}, {Elsaesser},
  {Errando}, {Ferenc}, {Fonseca}, {Font}, {Garc{\'{\i}}a L{\'o}pez},
  {Garczarczyk}, {Gaug}, {Godinovic}, {Hadasch}, {Herrero}, {Hildebrand},
  {H{\"o}hne-M{\"o}nch}, {Hose}, {Hrupec}, {Hsu}, {Jogler}, {Klepser},
  {Kr{\"a}henb{\"u}hl}, {Kranich}, {La Barbera}, {Laille}, {Leonardo},
  {Lindfors}, {Lombardi}, {Longo}, {L{\'o}pez}, {Lorenz}, {Majumdar}, {Maneva},
  {Mankuzhiyil}, {Mannheim}, {Maraschi}, {Mariotti}, {Mart{\'{\i}}nez},
  {Mazin}, {Meucci}, {Miranda}, {Mirzoyan}, {Miyamoto}, {Mold{\'o}n}, {Moles},
  {Moralejo}, {Nieto}, {Nilsson}, {Ninkovic}, {Orito}, {Oya}, {Paiano},
  {Paoletti}, {Paredes}, {Partini}, {Pasanen}, {Pascoli}, {Pauss}, {Pegna},
  {Perez-Torres}, {Persic}, {Peruzzo}, {Prada}, {Prandini}, {Puchades},
  {Puljak}, {Reichardt}, {Rhode}, {Rib{\'o}}, {Rico}, {Rissi}, {R{\"u}gamer},
  {Saggion}, {Saito}, {Salvati}, {S{\'a}nchez-Conde}, {Satalecka}, {Scalzotto},
  {Scapin}, {Schultz}, {Schweizer}, {Shayduk}, {Shore}, {Sierpowska-Bartosik},
  {Sillanp{\"a}{\"a}}, {Sitarek}, {Sobczynska}, {Spanier}, {Spiro}, {Stamerra},
  {Steinke}, {Struebig}, {Suric}, {Takalo}, {Tavecchio}, {Temnikov}, {Terzic},
  {Tescaro}, {Teshima}, {Torres}, {Vankov}, {Wagner}, {Weitzel}, {Zabalza},
  {Zandanel}, {Zanin}, {Neronov}, \& {Semikoz}}]{REF:2010.MAGIC}
{Aleksi{\'c}}, J., {et~al.} 2010, \aap, 524, A77+

\bibitem[{{Ando} \& {Kusenko}(2010)}]{REF:2010.AndoKus}
{Ando}, S., \& {Kusenko}, A. 2010, \apjl, 722, L39

\bibitem[{{Atwood} {et~al.}(2007){Atwood}, {Bagagli}, {Baldini}, {Bellazzini},
  {Barbiellini}, {Belli}, {Borden}, {Brez}, {Brigida}, {Caliandro}, {Cecchi},
  {Cohen-Tanugi}, {de Angelis}, {Drell}, {Favuzzi}, {Fukazawa}, {Fusco},
  {Gargano}, {Germani}, {Giannitrapani}, {Giglietto}, {Giordano}, {Himel},
  {Hirayama}, {Johnson}, {Katagiri}, {Kataoka}, {Kawai}, {Kroeger}, {Kuss},
  {Latronico}, {Longo}, {Loparco}, {Lubrano}, {Massai}, {Mazziotta}, {Minuti},
  {Mizuno}, {Morselli}, {Nelson}, {Nordby}, {Ohsugi}, {Omodei}, {Ozaki},
  {Pepe}, {Rain{\`o}}, {Rando}, {Razzano}, {Rich}, {Sadrozinski}, {Scolieri},
  {Sgr{\`o}}, {Spandre}, {Spinelli}, {Sugizaki}, {Tajima}, {Takahashi},
  {Takahashi}, {Yoshida}, {Young}, \& {Ziegler}}]{REF:2007.TKRPaper}
{Atwood}, W.~B., {et~al.} 2007, Astroparticle Physics, 28, 422

\bibitem[{{Atwood} {et~al.}(2009){Atwood}, {Abdo}, {Ackermann}, {Althouse},
  {Anderson}, {Axelsson}, {Baldini}, {Ballet}, {Band}, {Barbiellini}, \&
  et~al.}]{REF:2009.LATPaper}
---. 2009, \apj, 697, 1071

\bibitem[{{Boinee} {et~al.}(2003){Boinee}, {Cabras}, {de Angelis}, {Favretto},
  {Frailis}, {Giannitrapani}, {Milotti}, {Longo}, {Brigida}, {Gargano},
  {Giglietto}, {Mazziotta}, {Cecchi}, {Lubrano}, {Pepe}, {Baldini},
  {Cohen-Tanugi}, {Kuss}, {Latronico}, {Omodei}, {Spandre}, {Bogart}, {Dubois},
  {Kamae}, {Rochester}, {Usher}, {Burnett}, \&
  {Robinson}}]{REF:2003sngh.conf..141B}
{Boinee}, P., {et~al.} 2003, in Science with the New Generation of High Energy
  Gamma-Ray Experiments : Between Astrophysics and Astroparticle Physics, ed.
  {S.~Ciprini, A.~de Angelis, P.~Lubrano, \& O.~Mansutti}, 141

\bibitem[{{D'Avezac} {et~al.}(2007){D'Avezac}, {Dubus}, \&
  {Giebels}}]{2007A&A...469..857D}
{D'Avezac}, P., {Dubus}, G., \& {Giebels}, B. 2007, \aap, 469, 857

\bibitem[{{Dermer} {et~al.}(2011){Dermer}, {Cavadini}, {Razzaque}, {Finke},
  {Chiang}, \& {Lott}}]{REF:2011.DermerHalo}
{Dermer}, C.~D., {Cavadini}, M., {Razzaque}, S., {Finke}, J.~D., {Chiang}, J.,
  \& {Lott}, B. 2011, \apjl, 733, L21+

\bibitem[{{Dolag} {et~al.}(2011){Dolag}, {Kachelriess}, {Ostapchenko}, \&
  {Tom{\`a}s}}]{2011ApJ...727L...4D}
{Dolag}, K., {Kachelriess}, M., {Ostapchenko}, S., \& {Tom{\`a}s}, R. 2011,
  \apjl, 727, L4+

\bibitem[{{Elyiv} {et~al.}(2009){Elyiv}, {Neronov}, \&
  {Semikoz}}]{2009PhRvD..80b3010E}
{Elyiv}, A., {Neronov}, A., \& {Semikoz}, D.~V. 2009, \prd, 80, 023010

\bibitem[{{Essey} {et~al.}(2010){Essey}, {Kalashev}, {Kusenko}, \&
  {Beacom}}]{2010PhRvL.104n1102E}
{Essey}, W., {Kalashev}, O.~E., {Kusenko}, A., \& {Beacom}, J.~F. 2010,
  Physical Review Letters, 104, 141102

\bibitem[{{Essey} \& {Kusenko}(2010)}]{2010APh....33...81E}
{Essey}, W., \& {Kusenko}, A. 2010, Astroparticle Physics, 33, 81

\bibitem[{{Gabici} \& {Aharonian}(2007)}]{2007Ap&SS.309..465G}
{Gabici}, S., \& {Aharonian}, F.~A. 2007, \apss, 309, 465

\bibitem[{{Geant4 Collaboration} {et~al.}(2003){Geant4 Collaboration},
  {Agostinelli}, {Allison}, {Amako}, {Apostolakis}, {Araujo}, {Arce}, {Asai},
  {Axen}, {Banerjee}, {Barrand}, {Behner}, {Bellagamba}, {Boudreau}, {Broglia},
  {Brunengo}, {Burkhardt}, {Chauvie}, {Chuma}, {Chytracek}, {Cooperman},
  {Cosmo}, {Degtyarenko}, {dell'Acqua}, {Depaola}, {Dietrich}, {Enami},
  {Feliciello}, {Ferguson}, {Fesefeldt}, {Folger}, {Foppiano}, {Forti},
  {Garelli}, {Giani}, {Giannitrapani}, {Gibin}, {G{\'o}mez Cadenas},
  {Gonz{\'a}lez}, {Gracia Abril}, {Greeniaus}, {Greiner}, {Grichine},
  {Grossheim}, {Guatelli}, {Gumplinger}, {Hamatsu}, {Hashimoto}, {Hasui},
  {Heikkinen}, {Howard}, {Ivanchenko}, {Johnson}, {Jones}, {Kallenbach},
  {Kanaya}, {Kawabata}, {Kawabata}, {Kawaguti}, {Kelner}, {Kent}, {Kimura},
  {Kodama}, {Kokoulin}, {Kossov}, {Kurashige}, {Lamanna}, {Lamp{\'e}n}, {Lara},
  {Lefebure}, {Lei}, {Liendl}, {Lockman}, {Longo}, {Magni}, {Maire},
  {Medernach}, {Minamimoto}, {Mora de Freitas}, {Morita}, {Murakami},
  {Nagamatu}, {Nartallo}, {Nieminen}, {Nishimura}, {Ohtsubo}, {Okamura},
  {O'Neale}, {Oohata}, {Paech}, {Perl}, {Pfeiffer}, {Pia}, {Ranjard}, {Rybin},
  {Sadilov}, {di Salvo}, {Santin}, {Sasaki}, {Savvas}, {Sawada}, {Scherer},
  {Sei}, {Sirotenko}, {Smith}, {Starkov}, {Stoecker}, {Sulkimo}, {Takahata},
  {Tanaka}, {Tcherniaev}, {Safai Tehrani}, {Tropeano}, {Truscott}, {Uno},
  {Urban}, {Urban}, {Verderi}, {Walkden}, {Wander}, {Weber}, {Wellisch},
  {Wenaus}, {Williams}, {Wright}, {Yamada}, {Yoshida}, \&
  {Zschiesche}}]{2003NIMPA.506..250G}
{Geant4 Collaboration} {et~al.} 2003, Nuclear Instruments and Methods in
  Physics Research A, 506, 250

\bibitem[{{Grove} \& {Johnson}(2010)}]{2010SPIE.7732E..16G}
{Grove}, J.~E., \& {Johnson}, W.~N. 2010, in Proc. SPIE, Vol. 7732, 77320J

\bibitem[{{Hobbs} {et~al.}(2006){Hobbs}, {Edwards}, \&
  {Manchester}}]{REF:2006.Hobbs}
{Hobbs}, G.~B., {Edwards}, R.~T., \& {Manchester}, R.~N. 2006, \mnras, 369, 655

\bibitem[{{King}(1962)}]{1962AJ.....67..471K}
{King}, I. 1962, \aj, 67, 471

\bibitem[{{Mattox} {et~al.}(1996){Mattox}, {Bertsch}, {Chiang}, {Dingus},
  {Digel}, {Esposito}, {Fierro}, {Hartman}, {Hunter}, {Kanbach}, {Kniffen},
  {Lin}, {Macomb}, {Mayer-Hasselwander}, {Michelson}, {von Montigny},
  {Mukherjee}, {Nolan}, {Ramanamurthy}, {Schneid}, {Sreekumar}, {Thompson}, \&
  {Willis}}]{1996ApJ...461..396M}
{Mattox}, J.~R., {et~al.} 1996, \apj, 461, 396

\bibitem[{{Murase} {et~al.}(2008){Murase}, {Takahashi}, {Inoue}, {Ichiki}, \&
  {Nagataki}}]{2008ApJ...686L..67M}
{Murase}, K., {Takahashi}, K., {Inoue}, S., {Ichiki}, K., \& {Nagataki}, S.
  2008, \apjl, 686, L67

\bibitem[{{Neronov} {et~al.}(2010){Neronov}, {Semikoz}, {Kachelriess},
  {Ostapchenko}, \& {Elyiv}}]{2010ApJ...719L.130N}
{Neronov}, A., {Semikoz}, D., {Kachelriess}, M., {Ostapchenko}, S., \& {Elyiv},
  A. 2010, \apjl, 719, L130

\bibitem[{{Neronov} \& {Semikoz}(2009)}]{2009PhRvD..80l3012N}
{Neronov}, A., \& {Semikoz}, D.~V. 2009, \prd, 80, 123012

\bibitem[{{Neronov} {et~al.}(2011){Neronov}, {Semikoz}, {Tinyakov}, \&
  {Tkachev}}]{REF:2011.Neronov1}
{Neronov}, A., {Semikoz}, D.~V., {Tinyakov}, P.~G., \& {Tkachev}, I.~I. 2011,
  \aap, 526, A90+

\bibitem[{{Neronov} \& {Vovk}(2010)}]{REF:2010.Neronov2}
{Neronov}, A., \& {Vovk}, I. 2010, Science, 328, 73

\bibitem[{{Nolan} {et~al.}(2012){Nolan}, {Abdo}, {Ackermann}, {Ajello},
  {Allafort}, {Antolini}, {Atwood}, {Axelsson}, {Baldini}, {Ballet}, \&
  et~al.}]{2012ApJS..199...31N}
{Nolan}, P.~L., {et~al.} 2012, \apjs, 199, 31

\bibitem[{{Perkins} \& {VERITAS Collaboration}(2010)}]{REF:2010HEAD...11.3318P}
{Perkins}, J.~S., \& {VERITAS Collaboration}. 2010, in Bulletin of the American
  Astronomical Society, Vol.~42, AAS/High Energy Astrophysics Division \#11,
  708

\bibitem[{{Plaga}(1995)}]{1995Natur.374..430P}
{Plaga}, R. 1995, \nat, 374, 430

\bibitem[{Protassov {et~al.}(2002)Protassov, van Dyk, Connors, Kashyap, \&
  Siemiginowska}]{REF:2002.Protassov}
Protassov, R., van Dyk, D.~A., Connors, A., Kashyap, V.~L., \& Siemiginowska,
  A. 2002, The Astrophysical Journal, 571, 545

\bibitem[{{Rando}(2009)}]{REF:2009.LATPerf}
{Rando}, R. 2009, arXiv:0907.0626

\bibitem[{{Tavecchio} {et~al.}(2010){Tavecchio}, {Ghisellini}, {Foschini},
  {Bonnoli}, {Ghirlanda}, \& {Coppi}}]{REF:2010.Tavecchio}
{Tavecchio}, F., {Ghisellini}, G., {Foschini}, L., {Bonnoli}, G., {Ghirlanda},
  G., \& {Coppi}, P. 2010, \mnras, 406, L70

\bibitem[{{Taylor} {et~al.}(2011){Taylor}, {Vovk}, \&
  {Neronov}}]{2011A&A...529A.144T}
{Taylor}, A.~M., {Vovk}, I., \& {Neronov}, A. 2011, \aap, 529, A144

\bibitem[{{Vovk} {et~al.}(2012){Vovk}, {Taylor}, {Semikoz}, \&
  {Neronov}}]{2012ApJ...747L..14V}
{Vovk}, I., {Taylor}, A.~M., {Semikoz}, D., \& {Neronov}, A. 2012, \apjl, 747,
  L14

\bibitem[{{Weltevrede} {et~al.}(2010){Weltevrede}, {Johnston}, {Manchester},
  {Bhat}, {Burgay}, {Champion}, {Hobbs}, {K{\i}z{\i}ltan}, {Keith}, {Possenti},
  {Reynolds}, \& {Watters}}]{REF:2010.Weltevrede}
{Weltevrede}, P., {et~al.} 2010, \pasa, 27, 64

\bibitem[{{Woo} {et~al.}(2005){Woo}, {Urry}, {van der Marel}, {Lira}, \&
  {Maza}}]{2005ApJ...631..762W}
{Woo}, J.-H., {Urry}, C.~M., {van der Marel}, R.~P., {Lira}, P., \& {Maza}, J.
  2005, \apj, 631, 762

\end{thebibliography}

\begin{deluxetable}{llrrrr}
\tabletypesize{\scriptsize}
  \tablecaption{List of AGN from 1FGL catalog used for calibration of the on-orbit PSF (\irf{P6\_V11}). 
}
\tablewidth{0pt}
\tablehead{\colhead{Source} & \colhead{Association} & \colhead {Photon Index} &
  \colhead{Flux\tablenotemark{a}} & \colhead{$\sqrt{TS}$\tablenotemark{b}} & \colhead{$z$\tablenotemark{c}} }
\startdata
1FGL J0033.5$-$1921 & RBS 76 & 1.89 & 2.8 & 16 & 0.61 \\
1FGL J0120.5$-$2700 & PKS 0118$-$272 & 1.99 & 3.7  & 20 & 0.557 \\
1FGL J0136.5+3905   & B3 0133+388 & 1.73 & 4.5  & 24 & --- \\
1FGL J0137.0+4751   & OC 457 & 2.34 & 9.7  & 30 & 0.859 \\
1FGL J0217.9+0144   & PKS 0215+015 & 2.18 & 6.0  & 25 & 1.715 \\
1FGL J0221.0+3555   & B2 0218+35 & 2.33 & 6.4  & 26 & 0.685 \\
1FGL J0238.6+1637   & PKS 0235+164 & 2.14 & 32.7 & 72 & 0.94 \\
1FGL J0303.5$-$2406 & PKS 0301$-$243 & 1.98 & 4.6  & 23 & 0.26 \\
1FGL J0319.7+4130   & NGC 1275 & 2.13 & 17.3 & 45 & 0.0176 \\
1FGL J0334.4$-$3727 & CRATES J0334$-$3725 & 2.10 & 2.8  & 13 & --- \\
1FGL J0423.2$-$0118 & PKS 0420$-$01 & 2.42 & 5.8  & 22 & 0.916 \\
1FGL J0428.6$-$3756 & PKS 0426$-$380 & 2.13 & 25.7 & 63 & 1.03 \\
1FGL J0433.5+2905   & CGRaBS J0433+2905 & 2.13 & 4.5  & 16 & 0.97 \\
1FGL J0442.7$-$0019 & PKS 0440$-$00 & 2.44 & 6.3  & 24 & 0.844 \\
1FGL J0449.5$-$4350 & PKS 0447$-$439 & 1.95 & 11.1 & 40 & 0.205 \\
1FGL J0457.0$-$2325 & PKS 0454$-$234 & 2.21 & 32.5 & 73 & 1.003 \\
1FGL J0507.9+6738   & 1ES 0502+675 & 1.75 & 2.3  & 17 & 0.416 \\
1FGL J0509.3+0540   & CGRaBS J0509+0541 & 2.16 & 3.9  & 18 & --- \\
1FGL J0538.8$-$4404 & PKS 0537$-$441 & 2.27 & 21.3 & 53 & 0.892 \\
1FGL J0630.9$-$2406 & CRATES J0630$-$2406 & 1.87 & 3.1  & 17 & 1.238 \\
1FGL J0700.4$-$6611 & PKS 0700$-$661 & 2.15 & 4.7  & 19 & --- \\
1FGL J0719.3+3306   & B2 0716+33 &  2.15 & 6.9  & 26 & 0.779 \\
1FGL J0730.3$-$1141 & PKS 0727$-$11 & 2.33 & 20.7 & 44 & 1.589 \\
1FGL J0738.2+1741   & PKS 0735+17 & 2.02 & 4.4  & 20 & 0.424 \\
1FGL J0808.2$-$0750 & PKS 0805$-$07 & 2.14 & 10.0 & 32 & 1.837 \\
1FGL J0818.2+4222   & B3 0814+425 & 2.15 & 8.7  & 32 & 0.53 \\
1FGL J0825.8$-$2230 & PKS 0823$-$223 & 2.14 & 5.3  & 22 & 0.91 \\
1FGL J0920.9+4441   & B3 0917+449 & 2.28 & 14.0 & 44 & 2.19 \\
1FGL J0957.7+5523   & 4C +55.17 & 2.05 & 10.5 & 40 & 0.896 \\
1FGL J1015.1+4927   & 1ES 1011+496 & 1.93 & 8.7  & 36 & 0.20 \\
1FGL J1058.4+0134   & PKS 1055+01 & 2.29 & 7.1  & 27 & 0.888 \\
1FGL J1058.6+5628   & CGRaBS J1058+5628 & 1.97 & 5.7  & 30 & 0.888 \\
1FGL J1104.4+3812   & Mkn 421 & 1.81 & 26.1 & 76 & 0.03  \\
1FGL J1159.4+2914   & 4C +29.45 & 2.37 & 5.5  & 23 & 0.729 \\
1FGL J1221.5+2814   & W Com & 2.06 & 6.9  & 29 & 0.102 \\
1FGL J1224.7+2121   & 4C +21.35 & 2.55 & 2.5  & 13 & 0.432 \\
1FGL J1246.7$-$2545 & PKS 1244$-$255 & 2.31 & 8.3  & 26 & 0.635 \\
1FGL J1248.2+5820   & CGRaBS J1248+5820 & 2.18 & 4.5  & 23 & --- \\
1FGL J1253.0+5301   & CRATES J1253+5301 & 2.14 & 3.2  & 16 & --- \\
1FGL J1256.2$-$0547 & 3C 279 & 2.32 & 32.4 & 72 & 0.536 \\
1FGL J1312.4+4827   & CGRaBS J1312+4828 &  2.34 & 1.5  & 9  & 0.501 \\
1FGL J1344.2$-$1723 & CGRaBS J1344$-$1723 & 2.11 & 6.0  & 21 & --- \\
1FGL J1426.9+2347   & PKS 1424+240 & 1.83 & 10.2 & 40 & --- \\
1FGL J1444.0$-$3906 & PKS 1440-389 & 1.83 & 3.5  & 16 & 0.0655\\
1FGL J1457.5$-$3540 & PKS 1454$-$354 & 2.27 & 16.9 & 40 & 1.424 \\
1FGL J1504.4+1029   & PKS 1502+106 & 2.22 & 67.0 & 113 & 1.839 \\
1FGL J1517.8$-$2423 & AP Lib & 2.10 & 5.6  & 21 & 0.048 \\
1FGL J1522.1+3143   & B2 1520+31 & 2.42 & 15.9 & 48 & 1.487 \\
1FGL J1542.9+6129   & CRATES J1542+6129 & 2.14 & 5.2  & 25 & --- \\
1FGL J1555.7+1111   & PG 1553+113 & 1.66 & 13.7 & 51 & --- \\
1FGL J1725.0+1151   & CGRaBS J1725+1152	& 1.89 & 3.4  & 16 & 0.018 \\
1FGL J1802.5$-$3939 & BZU J1802$-$3940 & 2.25 & 10.4 & 24 & 0.296 \\
1FGL J1903.0+5539   & CRATES J1903+5540 & 1.86 & 3.6  & 17 & --- \\
1FGL J1917.7$-$1922 & CGRaBS J1917$-$192 & 1.88 & 3.3  & 15 & 0.137 \\
1FGL J1923.5$-$2104 & OV $-$235 & 2.17 & 11.9 & 33 & 0.874 \\
1FGL J2000.0+6508   & 1ES 1959+650 & 2.10 & 6.0  & 26 & 0.047 \\
1FGL J2009.5$-$4849 & PKS 2005$-$489 & 1.90 & 5.0  & 21 & 0.071 \\
1FGL J2025.6$-$0735 & PKS 2023$-$07 & 2.35 & 12.5 & 36 & 1.338 \\
1FGL J2056.3$-$4714 & PKS 2052$-$47 & 2.54 & 4.8  & 18 & 1.491 \\
1FGL J2139.3$-$4235 & CRATES J2139$-$4235 & 2.08 & 8.3  & 29 & --- \\
1FGL J2158.8$-$3013 & PKS 2155$-$304 & 1.91 & 27.1 & 70 & 0.116 \\
1FGL J2202.8+4216   & BL Lac & 2.38 & 7.3  & 21 & 0.069 \\
1FGL J2203.5+1726   & PKS 2201+171 & 2.39 & 4.3  & 18 & 1.076 \\
1FGL J2253.9+1608   & 3C 454.3 & 2.47 & 46.2 & 85 & 0.859 \\
1FGL J2329.2$-$4954 & PKS 2326$-$502 & 2.42 & 4.2  & 17 & 0.518 \\
\enddata
\tablenotetext{a}{Photon flux between 1 and 100 GeV in units of
  $10^{-9} ~ \textrm{photons} ~ \textrm{cm}^{-2}\textrm{s}^{-1}$
  obtained by summing the 1FGL photon flux values in the three bands
  from 1 to 100 GeV.}  
\tablenotetext{b}{Sum of the 1FGL $TS$ values in the three bands
  between 1 and 100 GeV.}
\tablenotetext{c}{Redshifts for sources in 1LAC are taken from
  \citet{REF:2010ApJ...715..429A}.  Redshifts for sources not in 1LAC
  are taken from the NASA Extragalactic Database (NED).}
\label{TABLE::PSFAGN}
\end{deluxetable}
\begin{deluxetable}{rr|rr|rr|rr|rr}
\tabletypesize{\scriptsize}
\tablecaption{Statistics for the Vela and Geminga pulsars and the low and high redshift BL Lacs in the energy range $1000-3162$ MeV for the analysis in Section 4.2. The models for the on-pulse selection $\nu_i^{on}$ are displayed next to the number of counts $n_i^{on}$ in each angular bin. The BL Lac model counts $\nu_i^{agn}$ are fit for the null case ($N^{halo}=0$) in Equation \ref{eq:binnedlike}.}
\tablewidth{0pt}
\tablehead{
&&\multicolumn{2}{c}{Vela}&\multicolumn{2}{c}{Geminga}&\multicolumn{2}{c}{BL Lac ($z<0.5$)}&\multicolumn{2}{c}{BL Lac ($z>0.5$)}\\
Bin edges (deg) & $m_i$  & $\nu_i^{on}$ & $n_i^{on}$ & $\nu_i^{on}$ & $n_i^{on}$ &  $\nu_i^{agn}$ & $n_i^{agn}$ &  $\nu_i^{agn}$ & $n_i^{agn}$}
\startdata
$0.000-0.083$ & 0.083 & $996.3$ & 955 & $516.7$ & 560 & 442 & 450.8 &203 &214.0\\
$0.083-0.124$ & 0.083 & $999.9$ & 988 & $512.7$ & 525 & 478 & 463.9&206 &215.9\\
$0.124-0.160$ & 0.083 & $999.8$ & 996 & $519.0$ & 523 & 486 & 472.4&249 &222.7\\
$0.160-0.199$ & 0.084 & $1000.8$ & 1012 & $518.7$ & 507 & 533 & 493.8&258 &226.5\\
$0.199-0.239$ & 0.083 & $1000.1$ & 981 & $512.2$ & 532 & 514 & 502.5&218& 224.4\\
$0.239-0.283$ & 0.083 & $1000.3$ & 1008 & $518.0$ & 510 & 508 & 519.2&205& 227.0\\
$0.283-0.336$ & 0.084 &  $1004.3$ & 1026 & $510.3$ & 488& 601 & 570.9&255 &240.5\\
$0.336-0.406$ & 0.084 & $1006.0$ & 1034 & $513.9$ & 485&667& 648.3&250& 254.4\\
$0.406-0.493$ & 0.083 & $1004.6$ & 999 & $513.2$ & 519&634 &737.7&288 &278.3\\
$0.493-0.630$ & 0.083 & $1013.6$ & 995 & $513.0$ & 532&1088 &1076.9&317 &340.5\\
$0.630-0.875$ & 0.083 & $1034.3$ & 1052& $517.9$ & 500&1799 &1967.6&496 &524.7\\
$0.875-4.000$ & 0.082 & $1874.3$ & 1902 & $765.8$ & 743&64785 &64631.0&13303& 13279.2\\
\enddata
\label{TABLE::PULSARANG}
\end{deluxetable}
\begin{deluxetable}{rr|rr|rr|rr|rr}
\tabletypesize{\scriptsize}
\tablecaption{As in Table \ref{TABLE::PULSARANG} in the energy range $3162-10000$ MeV.}
\tablewidth{0pt}
\tablehead{
&&\multicolumn{2}{c}{Vela}&\multicolumn{2}{c}{Geminga}&\multicolumn{2}{c}{BL Lac ($z<0.5$)}&\multicolumn{2}{c}{BL Lac ($z>0.5$)}\\
Bin edges (deg) & $m_i$  & $\nu_i^{on}$ & $n_i^{on}$ & $\nu_i^{on}$ & $n_i^{on}$ &  $\nu_i^{agn}$ & $n_i^{agn}$ &  $\nu_i^{agn}$ & $n_i^{agn}$}
\startdata
$0.000-0.043$ & 0.083 & $177.6$ & 175 & $104.4$ & 107&156 &148.4&71 & 59.8\\
$0.043-0.063$ & 0.083 & $178.9$ & 182 & $104.1$ & 101&133 &141.5&62 & 58.7\\
$0.063-0.082$ & 0.084 & $179.0$ & 188 & $105.1$ & 96&145 &145.9&50 & 56.8\\
$0.082-0.099$ & 0.083 & $178.7$ & 196 & $104.4$ & 87&162 & 151.7&62 & 58.8\\
$0.099-0.121$ & 0.083 & $178.8$ & 167 & $104.1$ & 116&170 & 155.3&53 & 57.5\\
$0.121-0.142$ & 0.084 & $179.1$ & 175 & $105.8$ & 110&177 & 158.3&72 & 60.9\\
$0.142-0.166$ & 0.083 & $178.6$ & 184 & $104.4$ & 99&121 & 140.5&70 & 60.6\\
$0.166-0.196$ & 0.083 & $178.9$ & 179 & $106.1$ & 106&131 & 146.0&29 & 54.4\\
$0.196-0.237$ & 0.083 & $178.5$ & 176 & $106.4$ & 109&171 & 162.9&68 & 61.8\\
$0.237-0.292$ & 0.084 & $179.5$ & 167 & $105.3$ & 118&149 & 163.2&57 & 61.7\\
$0.292-0.408$ & 0.083 & $180.7$ & 176 & $108.1$ & 113&207 & 212.0&66 & 70.6\\
$0.408-4.000$ & 0.083 & $329.6$ & 336 & $144.9$ & 140&13821 & 13817.1&2753 & 2751.5\\
\enddata
\end{deluxetable}
\begin{deluxetable}{rr|rr|rr|rr|rr}
\tabletypesize{\scriptsize}
\tablecaption{As in Table \ref{TABLE::PULSARANG} in the energy range $10000-31623$ MeV.}
\tablewidth{0pt}
\tablehead{
&&\multicolumn{2}{c}{Vela}&\multicolumn{2}{c}{Geminga}&\multicolumn{2}{c}{BL Lac ($z<0.5$)}&\multicolumn{2}{c}{BL Lac ($z>0.5$)}\\
Bin edges (deg) & $m_i$  & $\nu_i^{on}$ & $n_i^{on}$ & $\nu_i^{on}$ & $n_i^{on}$ &  $\nu_i^{agn}$ & $n_i^{agn}$ &  $\nu_i^{agn}$ & $n_i^{agn}$}
\startdata
$0.000-0.026$ & 0.078 & $12.1$ & 12 & $3.9$ & 4&53 & 51.6&9 & 10.3\\
$0.026-0.040$ & 0.088 & $13.6$ & 13 & $4.4$ & 5&61 & 59.0&8 & 10.7\\
$0.040-0.051$ & 0.088 & $13.6$ & 11 & $4.4$ & 7&55 & 54.6&17 & 14.3\\
$0.051-0.067$ & 0.088 & $13.6$ & 10 & $4.4$ & 8&62 & 59.8&12 & 12.3\\
$0.067-0.077$ & 0.083 & $12.9$ & 13 & $4.1$ & 4&33 & 37.4&16 & 13.5\\
$0.077-0.086$ & 0.088 & $13.6$ & 16 & $4.4$ & 2&44 & 46.4&11 & 11.9\\
$0.086-0.104$ & 0.088 & $13.6$ & 14 & $4.4$ & 4&57 & 56.2&11 & 11.9\\
$0.104-0.134$ & 0.088 & $13.6$ & 13 & $4.4$ & 5&72 & 67.5&23 & 16.9\\
$0.134-0.164$ & 0.088 & $13.6$ & 17 & $4.4$ & 1&43 & 46.1&11 & 12.1\\
$0.164-0.191$ & 0.083 & $12.9$ & 13 & $4.1$ & 4&46 & 47.6&5 & 9.3\\
$0.191-0.260$ & 0.088 & $13.6$ & 14 & $4.4$ & 4&60 & 59.8&14 & 13.8\\
$0.260-4.000$ & 0.053 & $38.2$ & 42 & $8.7$ & 6&3195 & 3194.9&638 & 637.9\\
\enddata
\label{TABLE::PULSARANG2}
\end{deluxetable}
\begin{deluxetable}{rr|rrrr}
\tabletypesize{\scriptsize}
\tablecaption{Statistics for the low-redshift BL Lacs in the energy range $31623-100000$ MeV for the analysis in Section 4.3. The models for the BL~Lacs, $\nu_i^{agn}$, is displayed next to the number of counts, $n$, in each angular bin. $N^{agn}m_i$ is displayed to highlight the equal statistics of the BL~Lacs in each angular bin.}
\tablewidth{0pt}
\tablehead{
&&\multicolumn{4}{c}{BL Lac ($z<0.5$)} \\
Bin edges (deg) & $m_i$ & $N^{agn}m_i$ & $N^{iso}b_i$ & $\nu_i^{agn}$ & $n_i^{agn}$}
\startdata
$0.000-0.019$ & 0.076 & $20.0$ & $0.0$ & $20.0$ & 20\\
$0.019-0.026$ & 0.080 & $21.0$ & $0.0$ & $21.0$ & 21\\
$0.026-0.033$ & 0.080 & $21.0$ & $0.0$ & $21.0$ & 21\\
$0.033-0.043$ & 0.084 & $22.0$ & $0.0$ & $22.0$ & 22\\
$0.043-0.049$ & 0.080 & $21.0$ & $0.0$ & $21.0$ & 21\\
$0.049-0.058$ & 0.080 & $21.0$ & $0.0$ & $21.0$ & 21\\
$0.058-0.073$ & 0.080 & $20.9$ & $0.1$ & $21.0$ & 21\\
$0.073-0.102$ & 0.083 & $21.7$ & $0.3$ & $22.0$ & 22\\
$0.102-0.130$ & 0.079 & $20.7$ & $0.3$ & $21.0$ & 21\\
$0.130-0.849$ & 0.161 & $42.0$ & $36.0$ & $78.0$ & 78\\
$0.849-2.739$ & 0.058 & $15.3$ & $346.7$ & $362.0$ & 362\\
$2.739-4.000$ & 0.058 & $15.1$ & $433.9$ & $449.0$ & 449\\
\enddata
\label{TABLE::AGNANG}
\end{deluxetable}

\begin{deluxetable}{lrrrrrr}
\tabletypesize{\scriptsize}

\tablecaption{Summary of the 95\% C.L. upper limits on the fraction ($f_{halo}=N^{halo}/(N^{halo}+N^{agn})$) of
  $\gamma$-ray emission from low-redshift BL Lacs attributable to a
  halo component for the PSF-convolved Disk and Gaussian halo models. }

\tablewidth{0pt}
\tablehead{ 
&\multicolumn{3}{c}{Disk}&\multicolumn{3}{c}{Gaussian}\\
\cline{2-4}\cline{5-7}
\colhead{Energy (MeV)} & 
\colhead{$\theta_{halo} = 0.1^{\circ}$} &  \colhead{$\theta_{halo} = 0.5^{\circ}$} &  \colhead{$\theta_{halo} = 1.0^{\circ}$} & 
\colhead{$\theta_{halo} = 0.1^{\circ}$} &  \colhead{$\theta_{halo} = 0.5^{\circ}$} &  \colhead{$\theta_{halo} = 1.0^{\circ}$}
}
\startdata
$1000-3162$ & 0.48 & 0.03 & 0.02 & 0.14 & 0.02 & 0.05 \\
$3162-10000$ & 0.25 & 0.05 & 0.16 & 0.09 & 0.11 & 0.29 \\
$10000-31623$ & 0.31 & 0.27 & 0.55 & 0.19 & 0.43 & 0.69 \\
\enddata

\label{TABLE::lzsum}

\end{deluxetable}








\begin{deluxetable}{lrrrrrr}
\tabletypesize{\scriptsize}

\tablecaption{Summary of the 95\% C.L. upper limits on the fraction
  ($f_{halo}=N^{halo}/(N^{halo}+N^{agn})$) of $\gamma$-ray emission
  from high-redshift BL Lacs attributable to a halo component for the
  PSF-convolved Disk and Gaussian halo models. }
\tablewidth{0pt}
\tablehead{ 
&\multicolumn{3}{c}{Disk}&\multicolumn{3}{c}{Gaussian}\\
\cline{2-4}\cline{5-7}
\colhead{Energy (MeV)} & 
\colhead{$\theta_{halo} = 0.1^{\circ}$} &  \colhead{$\theta_{halo} = 0.5^{\circ}$} &  \colhead{$\theta_{halo} = 1.0^{\circ}$} & 
\colhead{$\theta_{halo} = 0.1^{\circ}$} &  \colhead{$\theta_{halo} = 0.5^{\circ}$} &  \colhead{$\theta_{halo} = 1.0^{\circ}$}
}

\startdata
$1000-3162$ & 0.57 & 0.05 & 0.04 & 0.23 & 0.05 & 0.09 \\
$3162-10000$ & 0.25 & 0.07 & 0.20 & 0.12 & 0.15 & 0.35 \\
$10000-31623$ & 0.56 & 0.34 & 0.61 & 0.28 &0.50 & 0.73 \\
\enddata

\label{TABLE::hzsum}

\end{deluxetable}

\end{document}